\documentclass[%
nofootinbib,
 amsmath,amssymb,
 aps,
 prd,
twocolumn, superscriptaddress
]{revtex4-1}

\usepackage{float}
\usepackage[normalem]{ulem}

\usepackage{graphicx}
\usepackage{dcolumn}
\usepackage{bm}
\usepackage{hyperref}

\usepackage{xcolor}

\usepackage{color}

\begin{document}

\preprint{APS/123-QED} 

\title{Radial pulsations, moment of inertia and tidal deformability of dark energy stars }

\author{Juan M. Z. Pretel}
 \email{juanzarate@cbpf.br}
 \affiliation{
 Centro Brasileiro de Pesquisas F{\'i}sicas, Rua Dr.~Xavier Sigaud, 150 URCA, Rio de Janeiro CEP 22290-180, RJ, Brazil
}

\date{\today}

\begin{abstract}
We construct dark energy stars with Chaplygin-type equation of state (EoS) in the presence of anisotropic pressure within the framework of Einstein gravity. From the classification established by Iyer \textit{et al.}~[\href{https://iopscience.iop.org/article/10.1088/0264-9381/2/2/013}{Class.~Quantum~Grav.~\textbf{2}, 219 (1985)}], we discuss the possible existence of isotropic dark energy stars as compact objects. However, there is the possibility of constructing ultra-compact stars for sufficiently large anisotropies. We investigate the stellar stability against radial oscillations, and we also determine the moment of inertia and tidal deformability of these stars. We find that the usual static criterion for radial stability $dM/d\rho_c >0$ still holds for dark energy stars since the squared frequency of the fundamental pulsation mode vanishes at the critical central density corresponding to the maximum-mass configuration. The dependence of the tidal Love number on the anisotropy parameter $\alpha$ is also examined. We show that the surface gravitational redshift, moment of inertia and dimensionless tidal deformability undergo significant changes due to anisotropic pressure, primarily in the high-mass region. Furthermore, in light of the detection of gravitational waves GW190814, we explore the possibility of describing the secondary component of such event as a stable dark energy star in the presence of anisotropy.

\end{abstract}

\maketitle


\section{Introduction}

Different types of observations (such as Type Ia supernovae, structure formation and CMB anisotropies) indicate that our Universe is not only expanding, it is accelerating. Within the standard $\Lambda$CDM model (which is based on cold dark matter and cosmological constant in Einstein gravity), this cosmic acceleration is due to a smooth component with large negative pressure and repulsive gravity, the so-called dark energy. Such a model gives a good agreement with the recent observational data \cite{Aghanim2020}, but suffers from the well-known coincidence problem and the fine-tuning problem \cite{Weinberg1989, Padmanabhan2003}. The exact physical nature of dark energy is still a mystery and, consequently, the possibility that dark matter and dark energy could be different manifestations of a single substance has been considered \cite{KAMENSHCHIK2001265, Bento2002, Reis2003, Xu2012}. In that regard, it was shown that the inhomogeneous Chaplygin gas offers a simple unified model of dark matter and dark energy \citep{BILIC2002}. It was also argued that if the Universe is dominated by the Chaplygin gas a cosmological constant would be ruled out with high confidence \cite{Makler2003}.

Using the \textit{Planck} 2015 CMB anisotropy, type-Ia supernovae and observed Hubble parameter data sets, the full parameter space of the modified Chaplygin gas was measured by Li \textit{et al.}~\cite{Li2019}. Based on recent observations of high-redshift quasars, Zheng and colleagues \cite{Zheng2022} investigated a series of Chaplygin gas models as candidates for dark matter-energy unification. The application of the Hamilton-Jacobi formalism for generalized Chaplygin gas models was carried out in Ref.~\cite{Ignatov2021}. Additionally, it is worth mentioning that Odintsov \textit{et al.}~\cite{OdintsovGS2020} considered two different equations of state for dark energy (i.e., power-law and logarithmic effective corrections to the pressure). They showed that the power-law model only yielded some modest results, achieved under negative values of bulk viscosity, while the logarithmic scenario provide good fits in comparison to the $\Lambda$CDM model.

Another way to give rise to an accelerated expansion of the Universe is by modifying the geometry itself \cite{Copeland2006, Koyama2016}, namely, considering higher curvature corrections to the standard Einstein-Hilbert action. Under this outlook, the cosmic acceleration can be modeled in the scope of a scalar-tensor gravity theory \cite{Boisseau2000, Esposito2001}. Moreover, within the context of the so-called $f(R)$ theories \cite{Sotiriou2010, Felice2010}, the quadratic term in the Ricci scalar $R$ leads to an inflationary solution in the early Universe \cite{Starobinsky1980}, although such a model does not provide a late-time accelerated expansion. Nevertheless, the late-time acceleration era can be realized by terms containing inverse powers of $R$ \cite{Carroll2004}, though it was shown that this is not compatible with the solar system experiments \cite{Chiba2003}. For a comprehensive study on the evolution of the early and present Universe in $f(R)$ modified gravity, we refer the reader to the review articles \cite{Nojiri2011, Clifton2012, Nojiri2017} and references contained therein. On the other hand, the astrophysical implications due to the $f(R)$ modified gravitational Lagrangian on compact stars have been intensively investigated in the past few years \cite{Folomeev2018, Olmo2020, Astashenok2020, Astashenok2021, Numajiri2022, Nobleson2022, Pretel2022, Pretel2022JCAP}.

According to the aforementioned works, different dark energy models have been proposed in order to explain the mechanisms that lead to the cosmic acceleration. Only about $4\%$ of the Universe is made of familiar atomic matter, $20\%$ dark matter, and it turns out that roughly $76\%$ of the Universe is dark energy \cite{Frieman2008}. Within the context of General Relativity, dark energy is an exotic negative pressure contribution that can lead to the observed accelerated expansion. In the absence of consensus regarding a theoretical description for the current accelerated expansion of the Universe, theorists have proposed using the Chaplygin gas as a useful phenomenological description \cite{KAMENSHCHIK2001265}. If dark energy is distributed anywhere permeating ordinary matter, then it could be present in the interior of a compact star. Therefore, the purpose of this manuscript is to investigate the possible existence of compact stars with dark energy by assuming a Chaplygin-type EoS. For such stars to exist in nature, they need to be stable under small radial perturbations.

Adopting a description of dark energy by means of a phantom (ghost) scalar field, Yazadjiev \cite{Yazadjiev2011} constructed a general class of exact interior solutions describing mixed relativistic stars containing both ordinary matter and dark energy. The energy conditions and gravitational wave echoes of such stars were recently analyzed in Ref.~\cite{Sakti2021}. Furthermore, the effect of the dynamical scalar field quintessence dark energy on neutron stars was investigated in \cite{Smerechynskyi2021}. Panotopoulos and collaborators \cite{Panotopoulos2021} studied slowly rotating dark energy stars made of isotropic matter using the Chaplygin EoS. Bhar \cite{Bhar2021} proposed a model for a dark energy star made of dark and ordinary matter in the Tolman–Kuchowicz spacetime geometry. For further stellar models with dark energy we also refer the reader to Refs.~\cite{Chan2009, Farook2010, Ghezzi2011, Bhar2018, Tello2020, Estevez2021, Veneroni2021, Grammenos2021, Haghani2022}.

In addition, anisotropy in compact stars may arise due to strong magnetic fields, pion condensation, phase transitions, mixture of two fluids, bosonic composition, rotation, etc. Thus, regardless of the specific source of the anisotropy, it is more natural to think of anisotropic fluids when studying compact stars at densities above nuclear saturation density. In that regard, the literature offers some physically motivated functional relations for the anisotropy, see for example Refs.~\citep{BowersLiang1974, CHEW1981, Horvat2011, Doneva2012, HerreraBarreto2013, Raposo2019, Pretel2020EPJC}. However, we must point out that these anisotropic models are based on general assumptions (or ansatzes) that do not directly relate to exotic modifications of matter or gravity. Indeed, it has been argued that the deformation near the maximum neutron-star mass comes from the anisotropic pressure within these stars, which is caused by the distortion of Fermi surface predicted by the equation of state of the models \cite{Rizaldy2019}. Becerra-Vergara \textit{et al.}~\cite{Becerra2019} showed that the contribution of the fourth order corrections parameter ($a_4$) of the QCD perturbation on the radial and tangential pressure generate significant effects on the mass-radius relation and the stability of quark stars. It has also been shown that the stellar structure equations in Eddington-inspired Born-Infeld theory with isotropic matter can be recast into GR with a modified (apparent) anisotropic matter \cite{Danarianto2019}.

Motivated by the several works already mentioned, we aim to discuss the impact of anisotropy on the macroscopic properties of dark energy stars with Chaplygin-like EoS. We will address the following questions: Do these stars belong to families of compact or ultra-compact stars? How does anisotropy affect the compactness and radial stability of dark energy stars satisfying the causality condition? In particular, by adopting the phenomenological ansatz proposed by Horvat \textit{et al.}~\cite{Horvat2011}, we determine the radius, mass, gravitational redshift, frequency of the fundamental oscillation mode, moment of inertia and the dimensionless tidal deformability of anisotropic dark energy stars. The isotropic solutions are recovered when the anisotropy parameter vanishes, i.e.~when $\alpha= 0$.

The organization of this paper is as follows: In Sec.~\ref{Sec2} we start with a brief overview of relativistic stellar structure, describing the basic equations for radial pulsations, moment of inertia and tidal deformability. We then introduce the Chaplygin-like EoS and discuss its relation to the cosmological context in Sec.~\ref{Sec3}, as well as we present the anisotropy profile. Section \ref{Sec4} provides a discussion of the numerical results for the different physical properties of dark energy stars. Finally, our conclusions are summarized in Sec.~\ref{Sec5}.


\section{Stellar structure equations}\label{Sec2}

In order to study the basic features of compact stars with dark energy, in this section we briefly summarize the stellar structure equations in Einstein gravity. In particular, we focus on hydrostatic equilibrium structure, radial pulsations, moment of inertia, and tidal deformability.

The theory of gravity to be used in this work is general relativity, where the Einstein field equations are given by
\begin{equation}\label{FieldEq}
    G_{\mu\nu} \equiv R_{\mu\nu} - \frac{1}{2}g_{\mu\nu}R = 8\pi T_{\mu\nu} ,
\end{equation}
with $G_{\mu\nu}$ being the Einstein tensor, $R_{\mu\nu}$ the Ricci tensor, $R$ denotes the scalar curvature, and $T_{\mu\nu}$ is the energy-momentum tensor. Since we are interested in isolated compact stars, we consider that the spacetime can be described by the spherically symmetric four-dimensional line element 
\begin{equation}\label{metric}
ds^2 = -e^{2\psi}dt^2 + e^{2\lambda} dr^2 + r^2(d\theta^2 + \sin^2\theta d\phi^2) . 
\end{equation}

In addition, we model the compact-star matter by an anisotropic perfect fluid, whose energy-momentum tensor is given by
\begin{equation}\label{EMTensor}
  T_{\mu\nu} = (\rho + p_t) u_\mu u_\nu + p_t g_{\mu\nu} - \sigma k_\mu k_\nu ,
\end{equation}
where $\rho$ is the energy density, $\sigma \equiv p_t - p_r$ the anisotropy factor, $p_r$ the radial pressure, $p_t$ the tangential pressure, $u^\mu$ the four-velocity of the fluid, and $k^\mu$ is a unit four-vector. These four-vectors must satisfy $u_\mu u^\mu = -1$, $k_\mu k^\mu = 1$ and $u_\mu k^\mu = 0$. Notice that the stellar fluid becomes isotropic when $\sigma =0$.

\subsection{TOV equations}

When the stellar fluid remains in hydrostatic equilibrium, neither metric nor thermodynamic quantities depend on the time coordinate. This allows us to write $u^\mu = e^{-\psi}\delta_0^\mu$ and $k^\mu = e^{-\lambda}\delta_1^\mu$. Accordingly, the hydrostatic equilibrium of an anisotropic compact star is governed by the TOV equations:
\begin{align}
    \frac{dm}{dr} &= 4\pi r^2 \rho ,  \label{TOV1}  \\
    \frac{dp_r}{dr} &= -(\rho + p_r)\left( \frac{m}{r^2} + 4\pi rp_r \right) \left( 1 - \frac{2m}{r} \right)^{-1} + \frac{2\sigma}{r},    \label{TOV2}  \\
    \frac{d\psi}{dr} &= -\frac{1}{\rho + p_r}\frac{dp_r}{dr} + \frac{2 \sigma}{r(\rho + p_r)} ,  \label{TOV3}
\end{align}
which are obtained from Eqs.~(\ref{FieldEq})-(\ref{EMTensor}) together with the conservation law $\nabla_\mu T_1^{\ \mu} = 0$. The metric function $\lambda(r)$ is determined from the relation $e^{-2\lambda} = 1 - 2m/r$, where $m(r)$ is the gravitational mass within a sphere of radius $r$. 

By supplying an EoS for the radial pressure in the form $p_r = p_r(\rho)$ and a defined anisotropy relation for $\sigma$, the system of differential equations (\ref{TOV1})-(\ref{TOV3}) is then numerically integrated from the center at $r=0$ to the surface of the star $r =R$ which correspond to a vanishing pressure. Therefore, the above equations will be solved under the requirement of the following boundary conditions
\begin{equation}\label{BCTOV}
\rho(0) = \rho_c,   \ \quad   m(0) = 0,   \ \quad   \psi(R) = \frac{1}{2}\ln\left[ 1 - \frac{2M}{R} \right] , \
\end{equation}
where $\rho_c$ is the central energy density, and $M \equiv m(R)$ is the total mass of the star calculated at its surface. The numerical solution of the TOV equations describes the equilibrium background and allow us to obtain the metric components and fluid variables.

\subsection{Radial oscillations}

A rigorous analysis of the radial stability of compact stars requires the calculation of the frequencies of normal vibration modes. Such frequencies can be found by considering small deviations from the hydrostatic equilibrium state but maintaining the spherical symmetry of the star. In the linear treatment, where all quadratic (or higher-order) or mixed terms in the perturbations are discarded, one assumes that all perturbations in physical quantities are arbitrarily small. The fluid element located at $r$ in the unperturbed configuration is displaced to radial coordinate $r+ \xi(t,r)$ in the perturbed configuration, where $\xi$ is the Lagrangian displacement. All perturbations have a harmonic time dependence of the form $\sim e^{i\nu t}$, where $\nu$ is the oscillation frequency to be determined. Consequently, defining $\zeta \equiv \xi/r$, the adiabatic\footnote{In the adiabatic theory, it is assumed that the fluid elements of the star neither gain nor lose heat during the oscillation.} radial pulsations of anisotropic compact stars are governed by the following differential equations \cite{Pretel2020EPJC}
\begin{align}
  \frac{d\zeta}{dr} =& -\frac{1}{r}\left( 3\zeta + \frac{\Delta p_r}{\gamma p_r} + \frac{2\sigma\zeta}{\rho + p_r} \right) + \frac{d\psi}{dr}\zeta ,  \label{ROEq1}   \\
  \frac{d(\Delta p_r)}{dr} =&\ \zeta\left\lbrace \nu^2e^{2(\lambda - \psi)}(\rho + p_r)r - 4\frac{dp_r}{dr} \right.  \nonumber  \\
  &\left. - 8\pi (\rho + p_r)e^{2\lambda}rp_r + r(\rho + p_r)\left(\frac{d\psi}{dr}\right)^2 \right.  \nonumber \\
  &\left. + 2\sigma\left(\frac{4}{r} + \frac{d\psi}{dr} \right) + 2\frac{d\sigma}{dr} \right\rbrace + 2\sigma\frac{d\zeta}{dr}   \nonumber \\
  & - \Delta p_r \left[ \frac{d\psi}{dr} + 4\pi(\rho + p_r)re^{2\lambda} \right] + \frac{2}{r}\delta\sigma ,  \label{ROEq2}  \quad
\end{align}
where $\Delta p_r$ is the Lagrangian perturbation of the radial pressure and $\gamma= (1+\rho/p_r)dp_r/d\rho$ is the adiabatic index at constant specific entropy.

The above first-order time-independent equations (\ref{ROEq1}) and (\ref{ROEq2}) require boundary conditions set at the center and surface of the star, similar to a vibrating string fixed at its ends. Since Eq.~(\ref{ROEq1}) has a singularity at the origin, the following condition must be required
\begin{equation}\label{BCRO1}
    \Delta p_r = -\frac{2\sigma\zeta}{\rho + p_r}\gamma p_r -3\gamma\zeta p_r  \qquad \  {\rm as}  \qquad \  r\rightarrow 0 ,
\end{equation}
while the Lagrangian perturbation of the radial pressure at the surface must satisfy
\begin{equation}\label{BCRO2}
    \Delta p_r = 0  \qquad \  {\rm as}  \qquad \  r\rightarrow R .
\end{equation}

\subsection{Moment of inertia}

Suppose a particle is dropped from rest at a great distance from a rotating star, then it would experience an ever increasing drag in the direction of rotation as it approaches the star. Based on this description, we introduce the angular velocity acquired by an observer falling freely from infinity, denoted by $\omega(r,\theta)$. Here we will calculate the moment of inertia of an anisotropic dark energy star under the slowly rotating approximation \citep{Hartle1967}. This means that when we consider rotational corrections only to first order in the angular velocity of the star $\Omega$, the line element (\ref{metric}) is replaced by its slowly rotating counterpart, namely
\begin{align}\label{RotMetrci}
  ds^2 =& -e^{2\psi(r)}dt^2 + e^{2\lambda(r)}dr^2 + r^2(d\theta^2 + \sin^2\theta d\phi^2)  \nonumber  \\
  & -2\omega(r,\theta)r^2\sin^2\theta dtd\phi ,
\end{align}
and following Ref.~\cite{Hartle1967}, it is pertinent to define the difference $\varpi \equiv \Omega- \omega$ as the coordinate angular velocity of the fluid element at ($r,\theta$) seen by the freely falling observer.

Keep in mind that $\Omega$ is the angular velocity of the stellar fluid as seen by an observer at rest at some spacetime point $(t,r,\theta,\phi)$, and hence the four-velocity up to linear terms in $\Omega$ can be written as $u^\mu= (e^{-\psi},0,0,\Omega e^{-\psi})$. To this order, the spherical symmetry is still preserved and it is possible to extend the validity of the TOV equations (\ref{TOV1})-(\ref{TOV3}). Nonetheless, the $03$-component of the field equations contributes an additional differential equation for angular velocity. By retaining only first-order terms in $\Omega$, such component becomes
\begin{align}\label{03CompEq}
    \frac{e^{\psi-\lambda}}{r^4}\frac{\partial}{\partial r}\left[ e^{-(\psi+\lambda)}r^4\frac{\partial\varpi}{\partial r} \right] &+ \frac{1}{r^2\sin^3\theta}\frac{\partial}{\partial\theta}\left[ \sin^3\theta\frac{\partial\varpi}{\partial\theta} \right]  \nonumber  \\
    &= 16\pi(\rho+ p_t)\varpi . 
\end{align}

As in the case of isotropic fluids, we follow the same treatment carried out by Hartle \cite{Hartle1967, Hartle1973} and we assume that $\varpi$ can be written as
\begin{equation}\label{ExpandEq}
    \varpi(r,\theta) = \sum_{l=1}^\infty \varpi_l(r)\left( \frac{-1}{\sin\theta}\frac{dP_l}{d\theta} \right) ,
\end{equation}
where $P_l$ are Legendre polynomials. Taking this expansion into account, Eq.~(\ref{03CompEq}) becomes 
\begin{align}\label{OmegaEq1}
    \frac{e^{\psi-\lambda}}{r^4}\frac{d}{dr}\left[ e^{-(\psi+\lambda)}r^4\frac{d\varpi_l}{dr} \right] &- \frac{l(l+1)-2}{r^2}\varpi_l  \nonumber  \\
    &= 16\pi(\rho+ p_t)\varpi_l . 
\end{align}

At a distance far away from the star, where $e^{-(\psi + \lambda)}$ becomes unity, the asymptotic solution of Eq.~(\ref{OmegaEq1}) takes the form $\varpi_l(r) \rightarrow a_1r^{-l-2} + a_2r^{l-1}$. If spacetime is to be flat at large $r$, then $\omega \rightarrow 2J/r^3$ (or equivalently, $\varpi \rightarrow \Omega - 2J/r^3$) for $r \rightarrow \infty$, where $J$ is the total angular momentum of the star \cite{Hartle1967, Glendenning}. Therefore, comparing this with the asymptotic behavior of $\varpi_l(r)$, we find that $l=1$. As a result, $\varpi$ is a function only of the radial coordinate, and Eq.~(\ref{OmegaEq1}) reduces to 
\begin{equation}\label{OmegaEq2}
    \frac{e^{\psi-\lambda}}{r^4}\frac{d}{dr}\left[ e^{-(\psi+\lambda)}r^4\frac{d\varpi}{dr} \right] = 16\pi(\rho+ p_t)\varpi ,
\end{equation}
which can be integrated to give
\begin{equation}\label{ArbitraryEq1}
    \left[ r^4\frac{d\varpi}{dr} \right]_R = 16\pi\int_0^R(\rho+ p_t)r^4e^{\lambda-\psi}\varpi dr . 
\end{equation}

In view of Eq.~(\ref{ArbitraryEq1}), we can obtain the angular momentum $J$ and hence the moment of inertia $I= J/\Omega$ of a slowly rotating anisotropic star:
\begin{equation}\label{MomInerEq}
    I = \frac{8\pi}{3}\int_0^R (\rho+ p_r+ \sigma)e^{\lambda-\psi}r^4 \left( \frac{\varpi}{\Omega} \right)dr ,
\end{equation}
which reduces to the expression given in Ref.~\cite{Glendenning} for isotropic compact stars when $\sigma =0$. For an arbitrary choice of the central value $\varpi(0)$, the appropriate boundary conditions for the differential equation (\ref{OmegaEq2}) come from the requirements of regularity at the center of the star and asymptotic flatness at infinity, namely
\begin{align}\label{BCMomIner}
    \left. \frac{d\varpi}{dr}\right\vert_{r=0} &= 0,  &  \lim_{r\rightarrow\infty}\varpi &= \Omega .
\end{align}

Once the solution for $\varpi(r)$ is found, we can then determine the moment of inertia through the integral (\ref{MomInerEq}). It is remarkable that the above expression for $I$ is referred to as the ``slowly rotating'' approximation because it was obtained to lowest order in the angular velocity $\Omega$ \cite{Glendenning}. This means that the stellar structure equations are still given by the TOV equations (\ref{TOV1})-(\ref{TOV3}).

\subsection{Tidal deformability}

It is well known that the tidal properties of neutron stars are measurable in gravitational waves emitted from the inspiral of a binary neutron-star coalescence \cite{Most2018, Chatziioannou2020}. In that regard, here we also study the dimensionless tidal deformability of individual dark energy stars. To do so, we follow the procedure carried out by Hinderer \textit{et al.}~\cite{Hinderer2008} (see also Refs.~\cite{Damour2009, Binnington2009, Postnikov2021, Chaves2019, Dietrich2021, Kumari2021} for additional results). The basic idea is as follows: In a binary system, the deformation of a compact star due to the tidal effect created by the companion star is characterized by the tidal deformability parameter $\bar{\lambda} = -Q_{ij}/\mathcal{E}_{ij}$, where $Q_{ij}$ is the induced quadrupole moment tensor and $\mathcal{E}_{ij}$ is the tidal field tensor \cite{Chaves2019}. Namely, the latter describes the tidal field from the spacetime curvature sourced by the distant companion.

The tidal parameter is related to the tidal Love number $k_2$ through the relation\footnote{It should be noted that the tidal deformability parameter is being denoted by $\bar{\lambda}$ in order not to be confused with the metric component $\lambda$.}
\begin{equation}\label{TidalDeforEq}
    \bar{\lambda} = \frac{2}{3}k_2 R^5 , 
\end{equation}
but it is common in the literature to define the dimensionless tidal deformability $\Lambda = \bar{\lambda}/M^5$, so in our results we will focus on $\Lambda$. The calculation of $\bar{\lambda}$ requires considering linear quadrupolar perturbations (due to the external tidal field) to the equilibrium configuration. Thus, the spacetime metric is given by $g_{\mu\nu} = g_{\mu\nu}^0 + h_{\mu\nu}$, where $g_{\mu\nu}^0$ describes the equilibrium configuration and $h_{\mu\nu}$ is a linearized metric perturbation. For static and even-parity perturbations in the Regge-Wheeler gauge \cite{Regge1957}, the perturbed metric can be written as \cite{Hinderer2008} 
\begin{align}
    h_{\mu\nu} =& \nonumber  \\
    &\hspace{-0.7cm} {\rm diag} \left[ -e^{2\psi(r)}H_0, e^{2\lambda(r)}H_2, r^2K, r^2\sin^2\theta K \right]Y_{2m}(\theta, \phi) ,
\end{align}
where $H_0$, $H_2$ and $K$ are functions of the radial coordinate, and $Y_{lm}$ are the spherical harmonics for $l=2$.

Since the perturbed energy-momentum tensor is given by $\delta T_\mu^\nu = {\rm diag} (-\delta\rho, \delta p_r, \delta p_t, \delta p_t)$, the linearized field equations imply that:
\[
\begin{cases}
   H_0 = -H_2 \equiv H & \text{from} \quad \delta G_2^2 - \delta G_3^3 =0, \\
   K' = 2H\psi' + H' & \text{from} \quad \delta G_1^2 =0,  \\
   \delta p_t = \frac{H}{8\pi r}e^{-2\lambda}(\lambda' + \psi')Y_{2m} & \text{from} \quad \delta G_2^2 = 8\pi\delta p_t.
\end{cases}
\]

In addition, from $\delta G_0^0 - \delta G_1^1 = -8\pi(\delta\rho + \delta p_t)$, we can obtain the following differential equation \cite{Biswas2019}
\begin{equation}
    H'' + \mathcal{P}H' + \mathcal{Q}H =0 , 
\end{equation}
or alternatively,
\begin{equation}\label{yEq}
    ry' = -y^2 + (1 - r\mathcal{P})y - r^2\mathcal{Q} ,
\end{equation}
where we have defined 
\begin{align}
    y &\equiv r\frac{H'}{H} ,  \\
    \mathcal{P} &\equiv \frac{2}{r} + e^{2\lambda}\left[ \frac{2m}{r^2} + 4\pi r(p_r - \rho) \right] ,  \\
    \mathcal{Q} &\equiv 4\pi e^{2\lambda}\left[ 4\rho + 8p_r + \frac{\rho+ p_r}{\mathcal{A}v_{sr}^2}(1+ v_{sr}^2) \right] - \frac{6e^{2\lambda}}{r^2} - 4\psi'^2 ,
\end{align}
with $\mathcal{A} \equiv dp_t/dp_r$ and $v_{sr}$ being the radial speed of sound. 

By matching the internal solution with the external solution of the perturbed variable $H$ at the surface of the star $r= R$, we obtain the tidal Love number \cite{Biswas2019}
\begin{align}\label{LoveNumEq}
    k_2 &= \frac{8}{5}(1- 2C)^2C^5 \left[ 2C(y_R -1) - y_R+ 2 \right]  \nonumber  \\
    &\times \left\lbrace 2C[ 4(y_R+ 1)C^4 + (6y_R- 4)C^3 \right.  \nonumber  \\
    &\left.+\ (26- 22y_R)C^2 + 3(5y_R -8)C - 3y_R+ 6 \right]   \nonumber  \\
    &\left.+\ 3(1-2C)^2\left[ 2C(y_R- 1)- y_R +2 \right]\log(1-2C) \right\rbrace^{-1} ,
\end{align}
where $C \equiv M/R$ is the compactness of the star, and $y_R \equiv y(R)$ is obtained by integrating equation (\ref{yEq}) from the origin up to the stellar surface.


\section{Equation of state and anisotropy model }\label{Sec3}

As it is well known, a possible alternative to the Phantom and Quintessence fields is the Chaplygin gas, where the EoS assumes the form $p_r = -B/\rho$, with $B$ being a positive constant (given in $\rm m^{-4}$ units). In fact, it was argued that such gas could provide a solution to unify the effects of dark matter in the early times and dark energy in late times \cite{KAMENSHCHIK2001265, Zheng2022}. Although the literature provides a more generalized version for such EoS in the context of the Friedmann-Lemaître-Robertson-Walker Universe \cite{Bento2002, Reis2003, Cunha2004, Gorini2008, Piattella2010, Xu2012, Salahedin2022, Marttens2022}, here we will use the simplest form plus a linear term corresponding to a barotropic fluid, namely 
\begin{equation}\label{EoS}
    p_r = A\rho - \frac{B}{\rho} ,
\end{equation}
where $A$ is a positive dimensionless constant. Our model is characterized by two free parameters $A$ and $B$. Nevertheless, we must emphasize here that Li \textit{et al.}~\cite{Li2019} considered an equation of state with three degrees of freedom, specifically $p= A\rho - B/\rho^\alpha$, where $\alpha$ is an extra parameter. They carried out a statistical treatment of astronomical data in order to constrain the parameter space. In the light of the Markov chain Monte Carlo method, they found that at $2\sigma$ level, $\alpha = -0.0156^{+0.0982+0.2346}_{-0.1380-0.2180}$ and $A= 0.0009^{+0.0018+0.0030}_{-0.0017-0.0030}$ from CMB$+$JLA$+$CC data sets. In other words, the constants $\alpha$ and $A$ are very close to zero and hence the nature of unified dark matter-energy model is very similar to the cosmological standard $\Lambda$CDM model.

On the other hand, at astrophysics level, compact stars obeying the EoS (\ref{EoS}) have been investigated by several authors, see for example Refs.~\cite{Panotopoulos2021, Farook2010, Bhar2018, Tello2020, Estevez2021}. In this work we will adopt values of $A$ and $B$ for which appreciable changes in the mass-radius diagram can be visualized in order to compare our theoretical results with observational measurements of massive pulsars.

In order to describe physically realistic compact stars, the causality condition must be respected throughout the interior region of the star. In other words, the speed of sound (defined by $v_s \equiv \sqrt{dp/d\rho}$) cannot be greater than the speed of light. Thus, in view of Eq.~(\ref{EoS}), we have 
\begin{equation}\label{SpeedSr}
    v_{sr}^2 \equiv \frac{dp_r}{d\rho} = A + \frac{B}{\rho^2} ,
\end{equation}
and since the radial pressure vanishes at the surface of the star, then $B= A\rho^2$. Thereby, the causality condition $v_{sr}^2(R) = 2A <1$ implies that $A < 0.5$.

Besides, it is more realistic to consider stellar models where there exists a tangential pressure as well as a radial one, since anisotropies arise at high densities, i.e.~above the nuclear saturation density as considered in this work. Although the literature offers different functional relations to model anisotropic pressures at very high densities inside compact stars \citep{BowersLiang1974, CHEW1981, Horvat2011, Doneva2012, HerreraBarreto2013, Raposo2019}, here we adopt the simplest model, which was proposed by Horvat and collaborators \cite{Horvat2011}
\begin{equation}\label{AniModel}
    \sigma = \alpha\left( \frac{2m}{r} \right)p_r = \alpha\left(1- e^{-2\lambda}\right)p_r ,
\end{equation}
where $\alpha$ is a dimensionless parameter that controls the amount of anisotropy within the stellar fluid. This parameter can assume positive or negative values of the order of unity, see Refs.~\cite{Folomeev2018, Pretel2022, Horvat2011, Doneva2012, Pretel2020EPJC, Silva2015, Yagi2015, Rahmansyah2020, Rahmansyah2021, PretelMPLA2022}. Notice that the isotropic solutions are recovered when the value of $\alpha$ vanishes. Specifically, the anisotropy ansatz (\ref{AniModel}) has two important characteristics:~(i) the fluid becomes isotropic at the center generating regular solutions and (ii) the effect of anisotropy vanishes in the hydrostatic equilibrium equation in the Newtonian limit. Unlike this profile, the effect of anisotropy does not vanish in the hydrostatic equilibrium equation in the non-relativistic regime for the Bowers-Liang model \cite{BowersLiang1974}, which could be an unphysical trait as argued in Ref.~\cite{Yagi2015}. For a broader discussion on the different ways of generating static spherically symmetric anisotropic fluid solutions, we refer the reader to the recent review article \cite{Kumar2022}.

Since the Eulerian perturbation for the metric potential $\lambda$ can be written as $\delta\lambda= -4\pi r(\rho+ p_r)e^{2\lambda}\xi$ \cite{Pretel2020EPJC}, then $\delta\sigma$ takes the form
\begin{equation}
    \delta\sigma = \alpha\left[ (1- e^{-2\lambda})\delta p_r - 8\pi p_r(\rho + p_r)r^2\zeta \right] ,
\end{equation}
where it should be noted that the relation between the Eulerian and Lagrangian perturbations for radial pressure is given by $\Delta p_r = \delta p_r + r\zeta p'_r$. The above expression will be substituted in Eq.~(\ref{ROEq2}) when we discuss later the radial pulsations in the stellar interior for at least some values of $\alpha$.

\section{Numerical results }\label{Sec4}

\subsection{Equilibrium configurations}

So far we do not know exactly whether the millisecond pulsars (observed in compact binaries from optical spectroscopic and photometric measurements) are hadronic, quark or hybrid stars. In fact, it has been theorized that cold quark matter might exist at the core of heavy neutron stars \cite{Annala2020}. Despite the precise measurements of masses \cite{Demorest:2010bx, Antoniadis:2013pzd, Cromartie2019} and radii \cite{Miller2019, Riley2019, Raaijmakers2019}, such constraints are still unable to distinguish the theoretical predictions coming from the different models for strange stars and (hybrid) neutron stars. This means that the dense matter EoS within compact stars still remains poorly understood. Furthermore, a realistic compact star possesses high magnetic fields and rotation properties, which significantly alter its internal structure. For comparison reasons, it is therefore common to use the observational mass-radius measurements (in view of the detection of gravitational waves and electromagnetic signals) on the mass-radius diagrams for any type of EoS even being of different microscopic compositions. In that perspective, our theoretical results will be compared with  observational measurements.

\begin{figure*} 
\includegraphics[width=8.4cm]{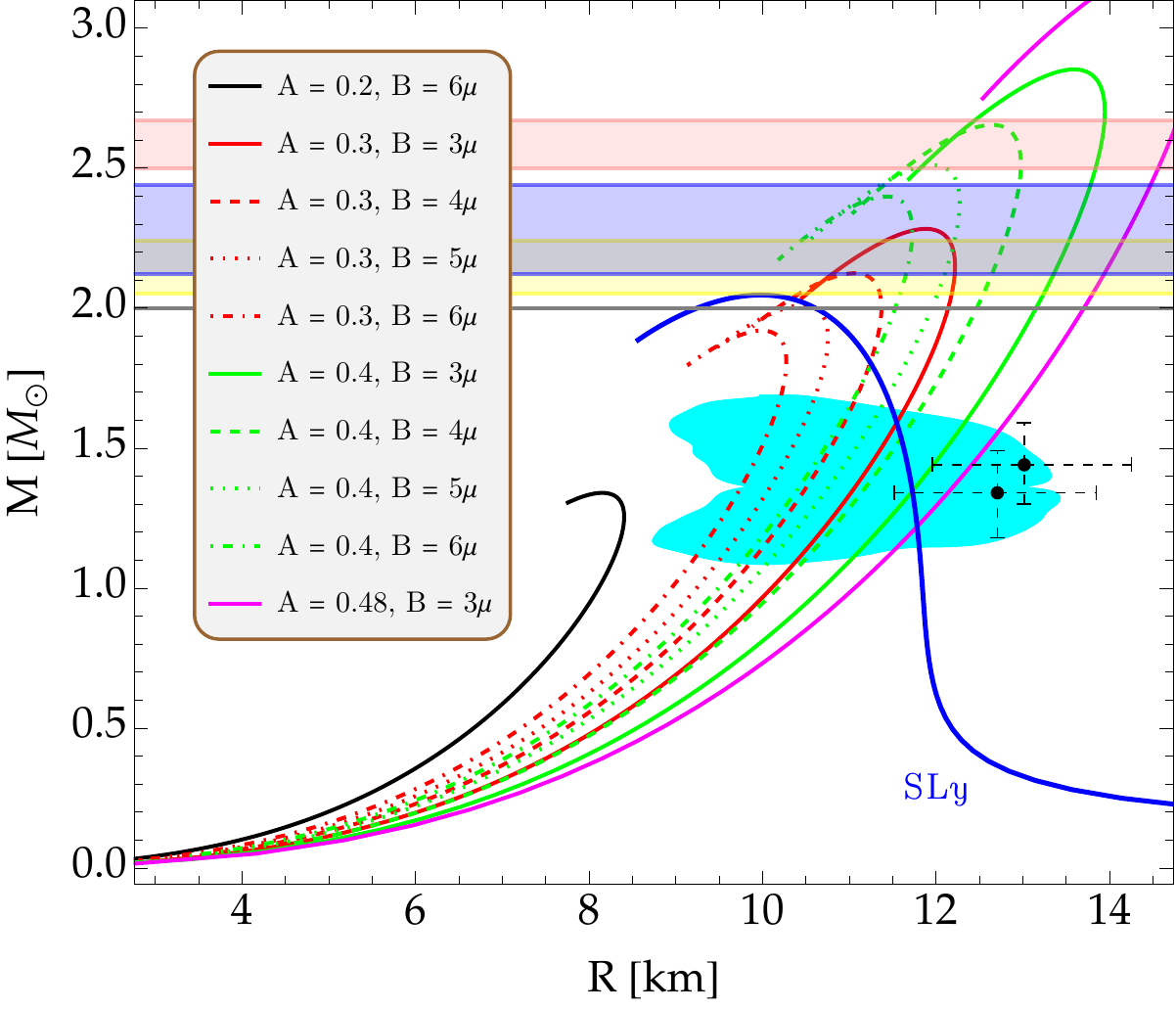}
\includegraphics[width=8.5cm]{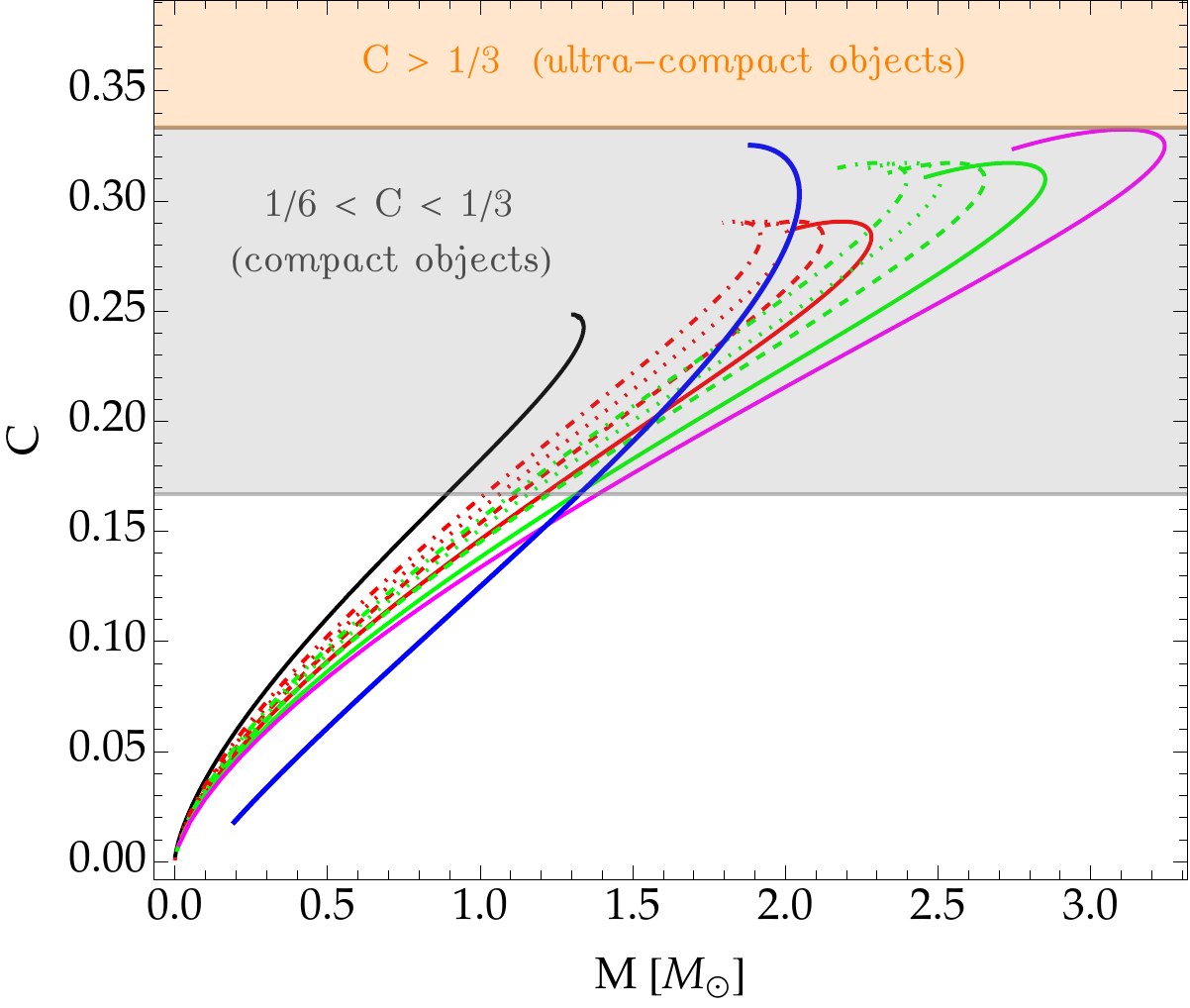}
  \caption{Left panel: Mass-radius diagrams for dark energy stars with Chaplygin-like EoS (\ref{EoS}) and isotropic pressure ($\sigma= 0$) for several values of the positive parameters $A$ and $B$. Here the constant $B$ is given in $\mu= 10^{-20}\rm \, m^{-4}$ units. The gray horizontal stripe at $2.0 M_\odot$ stands for the two massive NS pulsars J1614-2230 \cite{Demorest:2010bx} and J0348+0432 \cite{Antoniadis:2013pzd}. Yellow and blue regions represent the observational measurements of the masses of the highly massive NS pulsars J0740+6620 \cite{Cromartie2019} and J2215+5135 \cite{Linares2018}, respectively. The filled pink band stands for the lower mass of the compact object detected by the GW190814 event \citep{Abbott2020}, and the cyan area is the mass-radius constraint from the GW170817 event. Moreover, the NICER measurements for PSR J0030+0451 are displayed by black dots with their respective error bars \cite{Miller:2019cac, Riley:2019yda}. Right panel: Variation of the compactness with total gravitational mass, where the gray and orange stripes represent compact and ultra-compact objects, respectively, according to the classification given in Ref.~\cite{Iyer1985}. For comparison reasons, we have included the results corresponding to the SLy EoS \cite{DouchinHaensel2001} by blue curves in both plots.}
\label{figure1}
\end{figure*}

We begin our discussion of dark energy stars by considering the isotropic case (i.e., when $\sigma= 0$ in the TOV equations). We numerically integrate Eqs.~(\ref{TOV1})-(\ref{TOV3}) from the center up to the surface of the star through the boundary conditions (\ref{BCTOV}). As usual, the radius $R$ is determined when the pressure vanishes, and the total mass $M$ is calculated at the surface. The felt panel of Fig.~\ref{figure1} exhibits the mass-radius relations of dark energy stars for different values of parameters $A$ and $B$ in the EoS (\ref{EoS}). Remark that we have adopted values of $A$ less than $0.5$ in order to respect the causality condition. One can observe that small values of $A$ (see black curve) do not provide compact stars that fit current observational data. However, higher values of maximum mass can be obtained for larger values of $A$, see for example red and green curves. For a fixed value of $A$, the maximum mass decreases as the parameter $B$ increases. We perceive that the secondary component resulting from the gravitational-wave signal GW190814 \citep{Abbott2020} can be consistently described as a compact star with Chaplygin EoS (\ref{EoS}) for $A = 0.4$ and $B \in [4, 5]\mu$. Furthermore, the magenta curve fits very well with all observational data, but its maximum-mass value is above $3M_\odot$.

Another interesting feature of these stars is their compactness, defined by $C\equiv M/R$. According to the classification adopted by Iyer \textit{et al.}~\cite{Iyer1985}, the configurations shown in the mass-radius diagram correspond to compact stars, see the right plot of Fig.~\ref{figure1}. Besides, we can appreciate that the compactness of dark energy stars is of the order of the compactness of hadronic-matter stars, as is the case of the SLy EoS \cite{DouchinHaensel2001}, despite the fact that the maximum mass in the magenta configuration sequence can exceed $3M_\odot$. Nonetheless, as we will see later, the introduction of anisotropy can turn such stars into ultra-compact objects. Of course, this will depend on the amount of anisotropy in the stellar interior.

In order to include anisotropic pressures and investigate their effects on the internal structure of dark energy stars, we will adopt two specific models with the following parameters
\begin{itemize}
    \item[$\star$] \textbf{Model I:} $A = 0.3$, \  $B = 6.0\mu$ ,
    \item[$\star$] \textbf{Model II:} $A = 0.4$, \  $B = 5.2\mu$ ,
\end{itemize}
which are models favored by observational measurements according to the left panel of Fig.~\ref{figure1}. Moreover, model II precisely corresponds to the first model considered by Panotopoulos \textit{et al.}~\cite{Panotopoulos2021}.

\begin{figure*} 
\includegraphics[width=5.75cm]{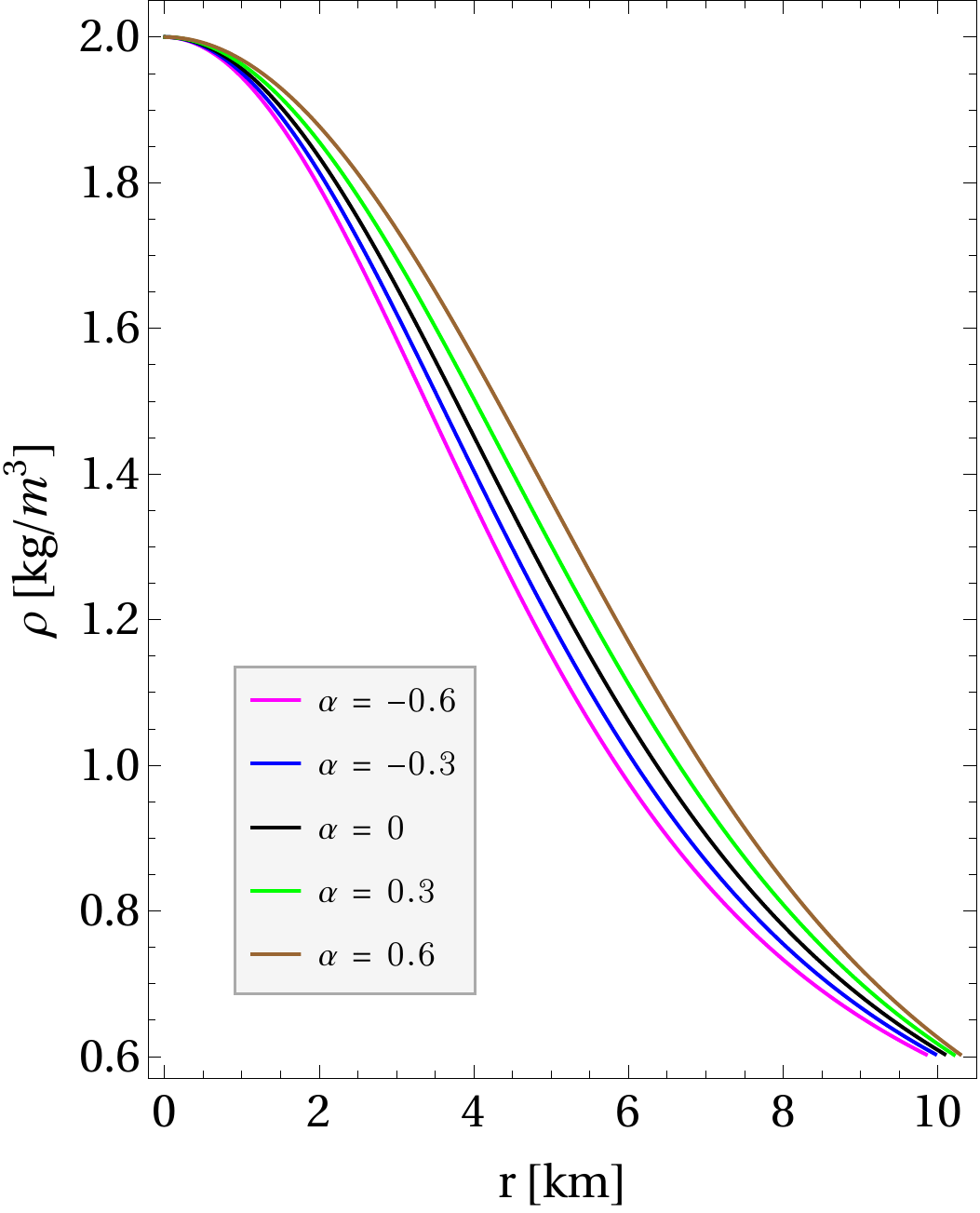}
\includegraphics[width=5.515cm]{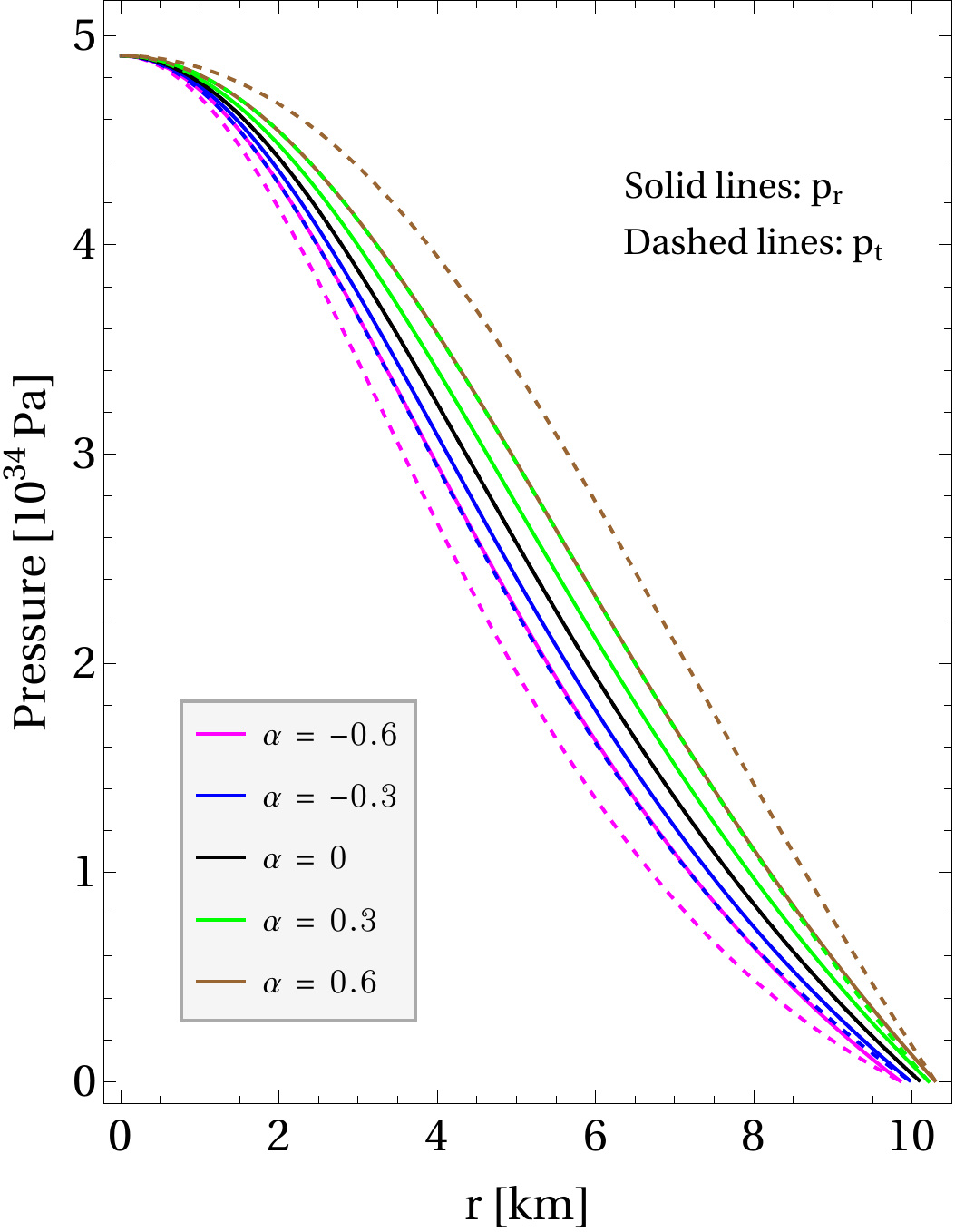}
\includegraphics[width=5.755cm]{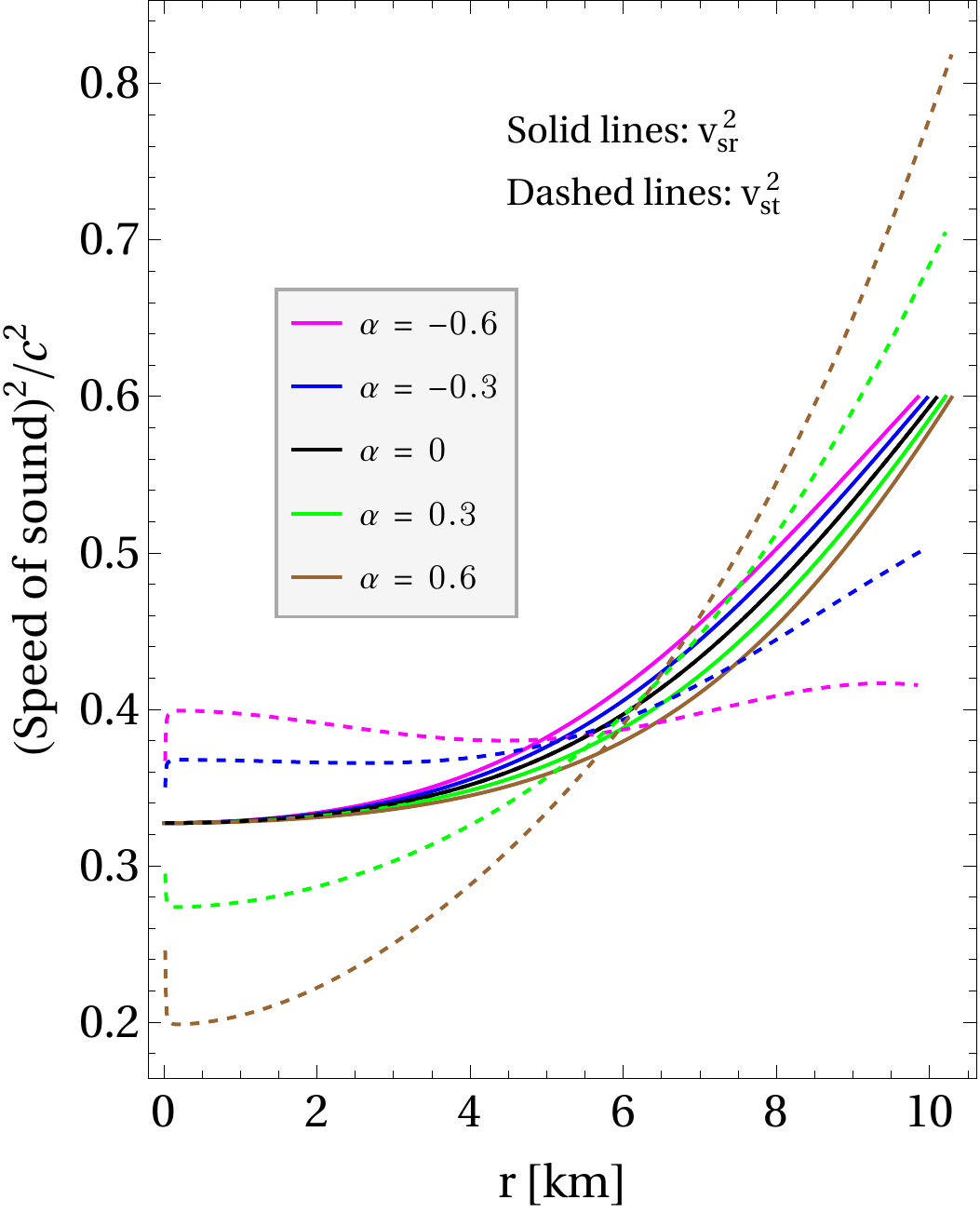}
  \caption{Radial behavior of the mass density (left panel), pressures (middle panel) and squared speed of sound (right panel) inside an anisotropic dark energy star with central density $\rho_c = 2.0 \times 10^{18}\, \rm kg/m^3$ and several values of the parameter $\alpha$. All plots correspond to model I and the black curves represent the isotropic solutions. Note that both the radial and tangential speed of sound obey the causality condition. Furthermore, one can observe that the increase in $\alpha$ leads to larger radii, and the anisotropy is more pronounced in the intermediate regions. }
\label{figure2}
\end{figure*}

Similar to the isotropic case, we numerically solve the hydrostatic background equations (\ref{TOV1})-(\ref{TOV3}) with boundary conditions (\ref{BCTOV}), but taking into account the anisotropy profile (\ref{AniModel}). For instance, for the model I and a central density $\rho_c = 2.0 \times 10^{18}\, \rm kg/m^3$, Fig.~\ref{figure2} illustrates the mass density, pressure and squared speed of sound as functions of the radial coordinate for different values of the free parameter $\alpha$. We can see that the internal structure of a dark energy star is affected by the presence of anisotropy. In effect, the radius of the star increases (decreases) for more positive (negative) values of $\alpha$. In addition, we remark that the speed of sound, both radial and tangential, respect the causality condition. This has also been verified for other values of central density considered in the construction of Fig.~\ref{figure1}.

\begin{figure*} 
\includegraphics[width=8.4cm]{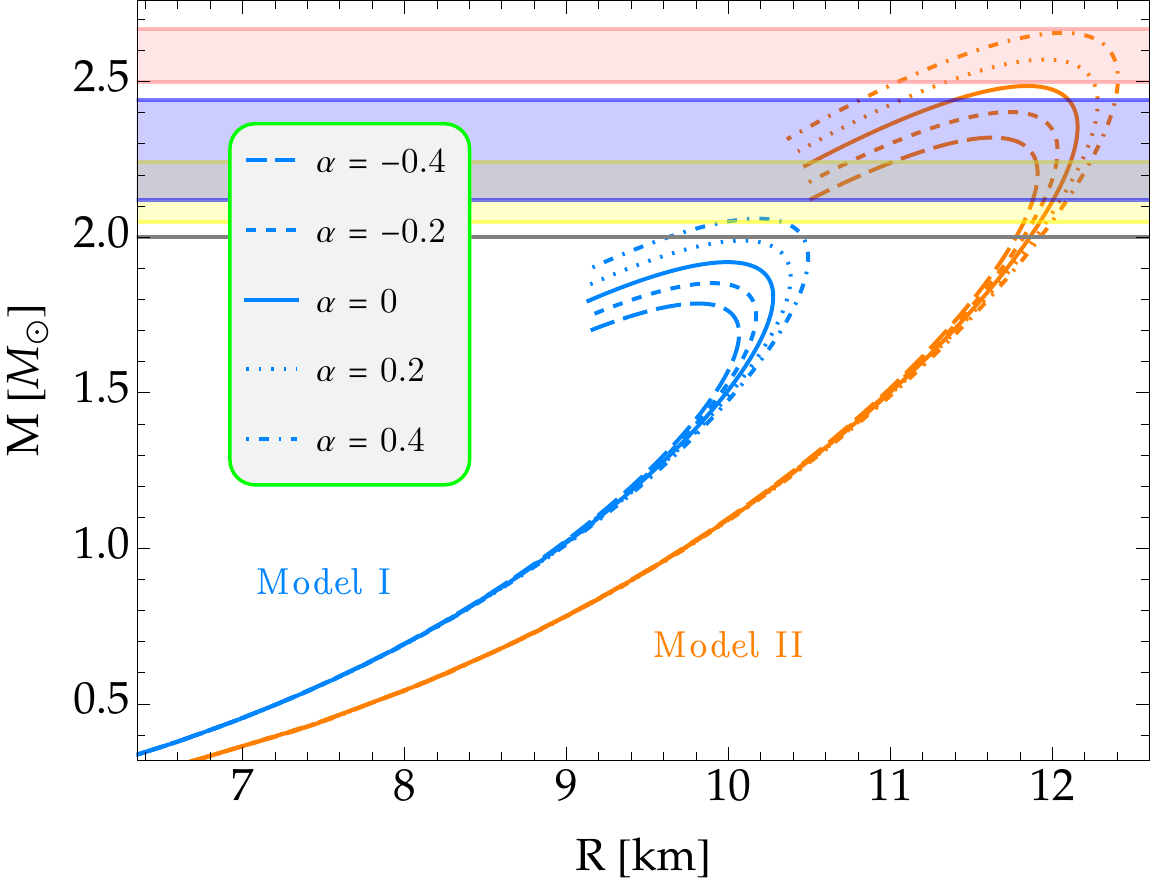}
\includegraphics[width=8.3cm]{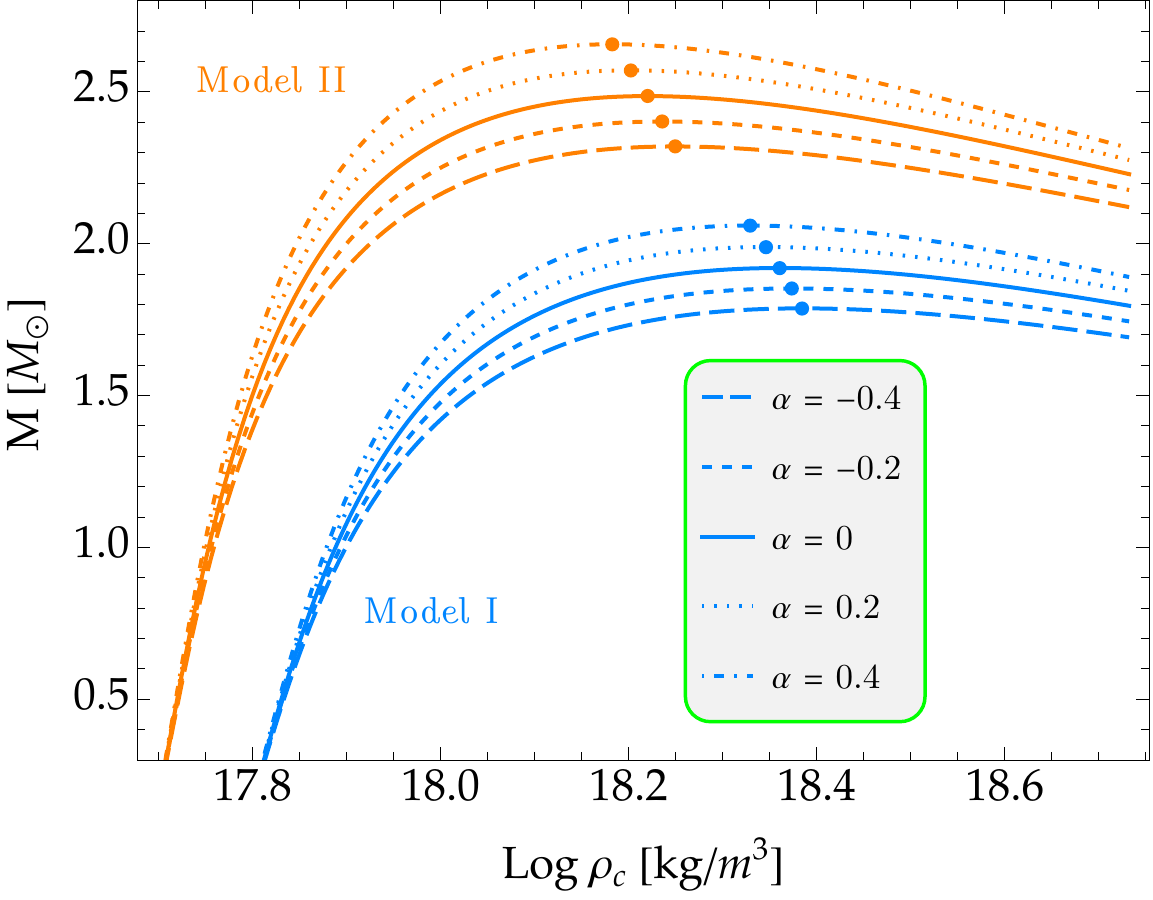}
  \caption{Mass-radius diagram (left panel) and mass-central density relation (right panel) for anisotropic dark energy stars as predicted by model I (blue curves) and II
  (orange curves) with anisotropy profile (\ref{AniModel}) for several values of $\alpha$. The colored bands in the left plot represent the same as in Fig.~\ref{figure1}. Moreover, the full blue and orange circles on the right plot indicate the maximum-mass points for model I and II, respectively. Note that the maximum-mass values for model II correspond to lower central densities than those for model I, however, model II allows larger masses (see also Table \ref{table1}). The critical central density corresponding to the maximum point on the $M(\rho_c)$ curve is modified by the presence of anisotropy for both models. }
\label{figure3}
\end{figure*}

Varying the central density, we obtain the mass-radius diagrams and mass-central density relations for models I and II, as shown in Fig.~\ref{figure3}. We observe that the substantial changes introduced by anisotropy in dark energy stars occur in the high-mass branch (close to the maximum-mass point), while the effects are irrelevant at low central densities. The maximum-mass values increase as the parameter $\alpha$ increases (see also the data in Table \ref{table1}). Note that model I without anisotropic pressures is not capable of generating maximum masses above $2M_\odot$. Nevertheless, the inclusion of anisotropies (see the blue curve for $\alpha = 0.4$) allows a significant increase in the maximum mass and hence a more favorable description of the compact objects observed in nature. On the other hand, model II with anisotropies (see orange curves) fits better with the observational measurements. In particular, in view of the lower mass of the compact object from the coalescence GW190814 \citep{Abbott2020}, two curves are particularly outstanding. In other words, such object can be well described as an anisotropic dark energy star when $\alpha= 0.2$ and $\alpha= 0.4$. Moreover, model II with negative anisotropies (such as $\alpha= -0.4$) favors the description of the massive pulsar J2215+5135 \citep{Linares2018}.

\begin{figure*} 
\includegraphics[width=8.4cm]{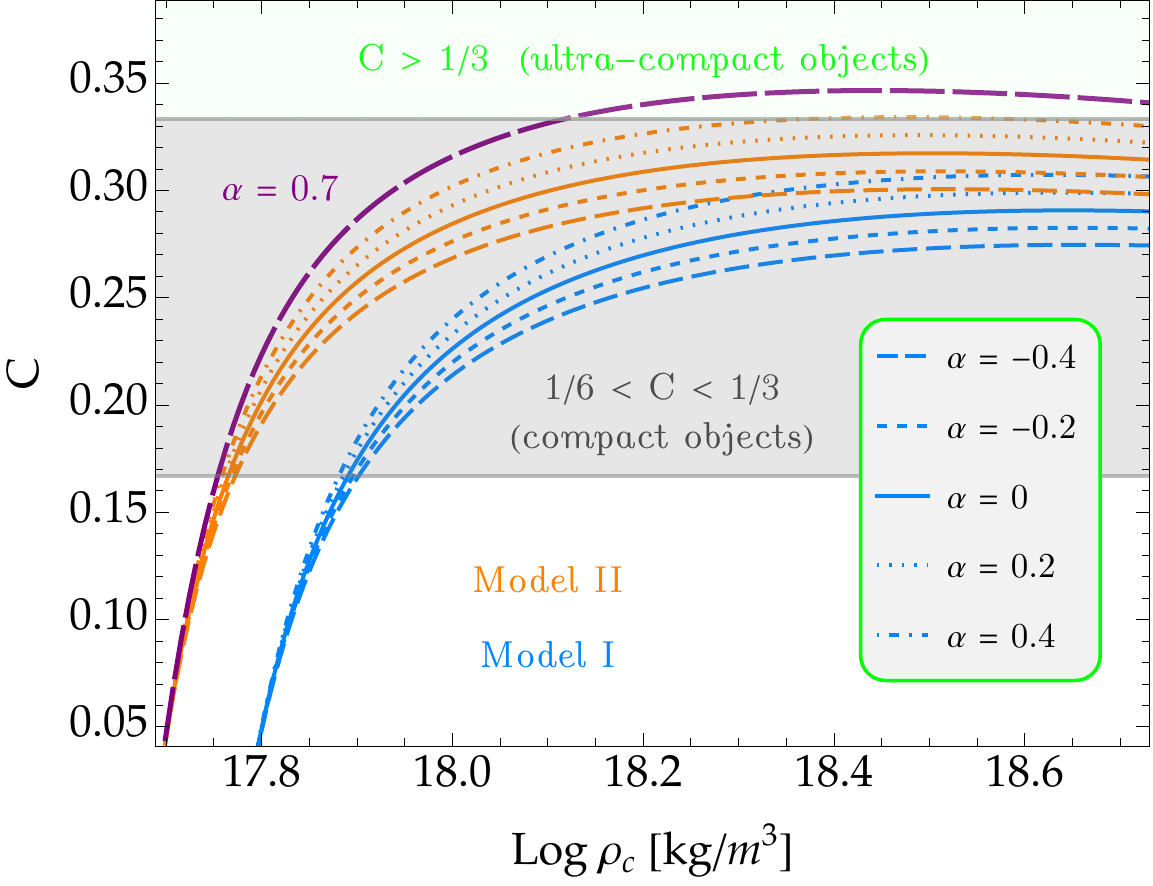}
\includegraphics[width=8.3cm]{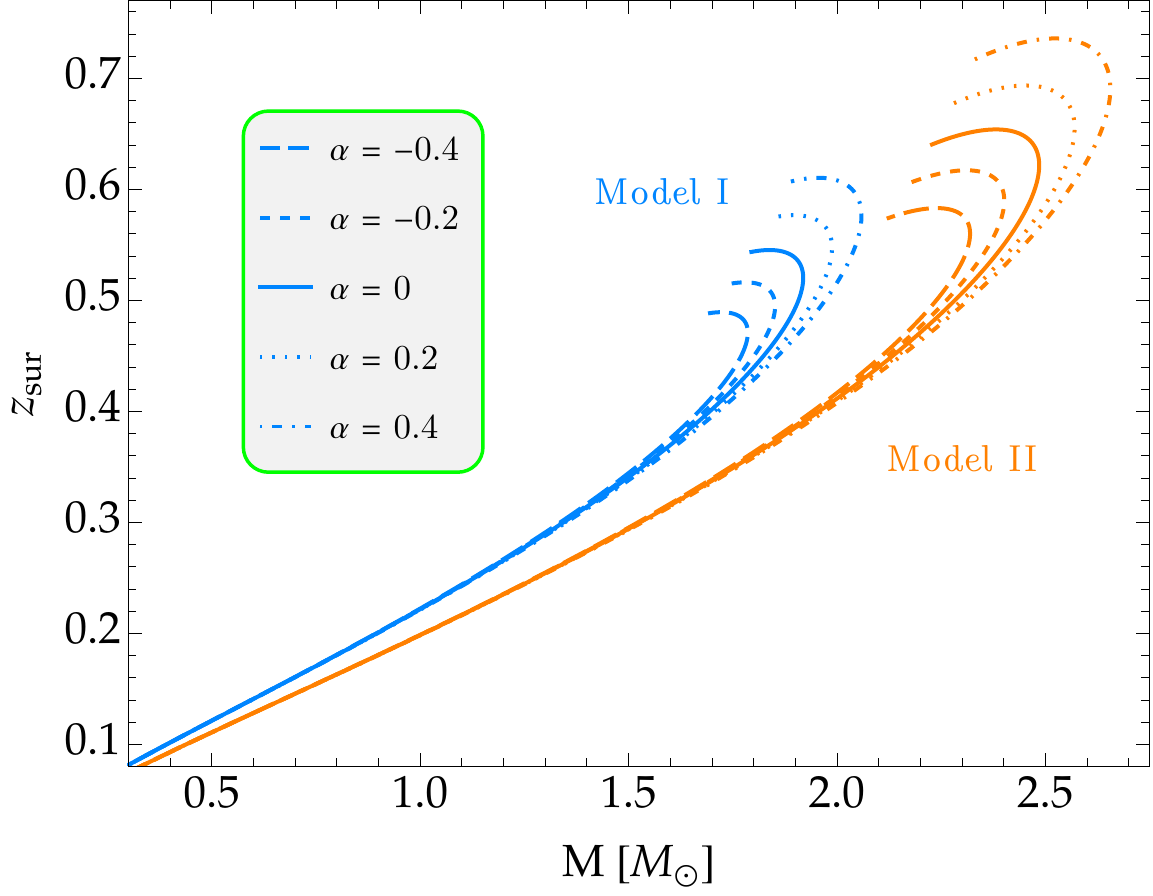}
  \caption{Left panel: Variation of the compactness with central density for several anisotropic dark energy star sequences. The gray and light-green stripes represent compact and ultra-compact objects, respectively, according to the classification established by Iyer \textit{et al.}~\cite{Iyer1985}. Positive anisotropy results in increased compactness for sufficiently high central densities, while the opposite occurs for negative anisotropy. Note also that dark energy stars would correspond to ultra-compact objects if $\alpha >0.4$ for model II, see for instance the purple curve for $\alpha= 0.7$. Right panel: Surface gravitational redshift as a function of the total mass. In the high-redshift region it can be observed that positive (negative) anisotropy increases (decreases) the value of $z_{\rm sur}$. Meanwhile, the effect of anisotropy is irrelevant for sufficiently low redshifts.}
\label{figure4}
\end{figure*}

The left panel of Fig.~\ref{figure4} describes the behavior of compactness as a function of central density. Positive anisotropies lead to an increase in compactness, mainly in the high-central-density branch. Remarkably, for sufficiently large values of $\alpha$ (see purple curve), it is possible to obtain anisotropic dark energy stars as ultra-compact objects.

The gravitational redshift, conventionally defined as the fractional change between observed and emitted wavelengths compared to emitted wavelength, in the case of a Schwarzschild star is given by \cite{Glendenning}
\begin{equation}\label{RedshiftEq}
    z_{\rm sur} = e^{\lambda(R)} -1 = \left( 1- \frac{2M}{R} \right)^{-1/2} -1 .
\end{equation}
In the right plot of Fig.~\ref{figure4}, the surface gravitational redshift is plotted as a function of the total mass for both models I and II. This plot indicates that the gravitational redshift of light emitted at the surface of a dark energy star is substantially affected by the anisotropy in the high-mass region, while the changes are negligible for sufficiently low masses. For a fixed value of central density, Table \ref{table2} shows that positive (negative) anisotropy increases (decreases) the value of the redshift.

\begin{table}
\caption{\label{table1} 
Maximum-mass configurations with Chaplygin-like EoS (\ref{EoS}) for model I and II. The energy density values correspond to the critical central density where the function $M(\rho_c)$ is a maximum on the right plot of Fig.~\ref{figure3}. }
\begin{ruledtabular}
\begin{tabular}{c|cccc}
Model  &  $\alpha$  &  $\rho_c$ [$10^{18}\, \rm kg/m^3$]  &  $R$ [\rm{km}]  &  $M$ [$M_\odot$]  \\
\colrule
  &  $-0.4$  &  2.424  &  9.812  &  1.786  \\
  &  $-0.2$  &  2.364  &  9.902  &  1.852  \\
I  &  $0$  &  2.295  &  9.994  &  1.919  \\
  &  $0.2$  &  2.219  &  10.086  &  1.988  \\
  &  $0.4$  &  2.135  &  10.180  &  2.059  \\
\colrule
  &  $-0.4$  &  1.777  &  11.630  &  2.320  \\
  &  $-0.2$  &  1.721  &  11.738  &  2.402  \\
II  &  $0$  &  1.661  &  11.845  &  2.486  \\
  &  $0.2$  &  1.594  &  11.955  &  2.570  \\
  &  $0.4$  &  1.523  &  12.065  &  2.565  \\
\end{tabular}
\end{ruledtabular}
\end{table}

\begin{table*}
\caption{\label{table2}
Radius, mass, redshift, fundamental mode frequency ($f_0= \nu_0/2\pi$), moment of inertia and dimensionless tidal deformability of dark energy stars with central energy density $\rho_c = 1.5 \times 10^{18}\, \rm kg/m^3$ as predicted by models I and II for several values of the anisotropy parameter $\alpha$. Remarkably, with the exception of the fundamental mode frequency and tidal deformability, these properties undergo a significant increase as $\alpha$ increases. }
\begin{ruledtabular}
\begin{tabular}{c|ccccccc}
Model  &  $\alpha$  &  $R$ [\rm{km}]  &  $M$ [$M_\odot$]   &   $z_{\rm sur}$   &  $f_0$ [kHz]   &  $I$ [$10^{38}\, \rm kg \cdot m^2$]   &  $\Lambda$  \\
\colrule
  & $-0.4$  &  10.062  &  1.713  &  0.418  &  2.414  &  1.695  &  13.278  \\
  & $-0.2$  &  10.163  &  1.781  &  0.440  &  2.312  &  1.820  &  10.709  \\
I  & $0$  &  10.263  &  1.852  &  0.463  &  2.201  &  1.957  &  8.598  \\
  & $0.2$  &  10.361  &  1.926  &  0.489  &  2.081  &  2.105  &  6.868  \\
  & $0.4$  &  10.456  &  2.003  &  0.518  &  1.950  &  2.265  &  5.454  \\
\colrule
  & $-0.4$  &  11.767  &  2.310  &  0.543  &  1.131  &  3.298  &  4.889  \\
  & $-0.2$  &  11.859  &  2.395  &  0.574  &  0.998  &  3.531  &  3.823  \\
II  & $0$  &  11.944  &  2.481  &  0.609  &  0.840  &  3.778  &  2.978  \\
  & $0.2$  &  12.019  &  2.569  &  0.647  &  0.637  &  4.037  &  2.309  \\
  & $0.4$  &  12.083  &  2.656  &  0.688  &  0.315  &  4.303  &  1.782  \\
\end{tabular}
\end{ruledtabular}
\end{table*}

\subsection{Oscillation spectrum }

A necessary condition (the well-known $M(\rho_c)$ method) for stellar stability is that stable stars must lie in the region where $dM/d\rho_c> 0$. According to the right plot of Fig.~\ref{figure3}, the full blue and orange circles on each curve indicate the onset of instability for each family of equilibrium solutions. However, a sufficient condition is to calculate the frequencies of the radial vibration modes for each central density \cite{Glendenning}. Here we will analyze if both methods are compatible in the case of dark energy stars including anisotropic pressure.

\begin{figure*} 
\includegraphics[width=8.6cm]{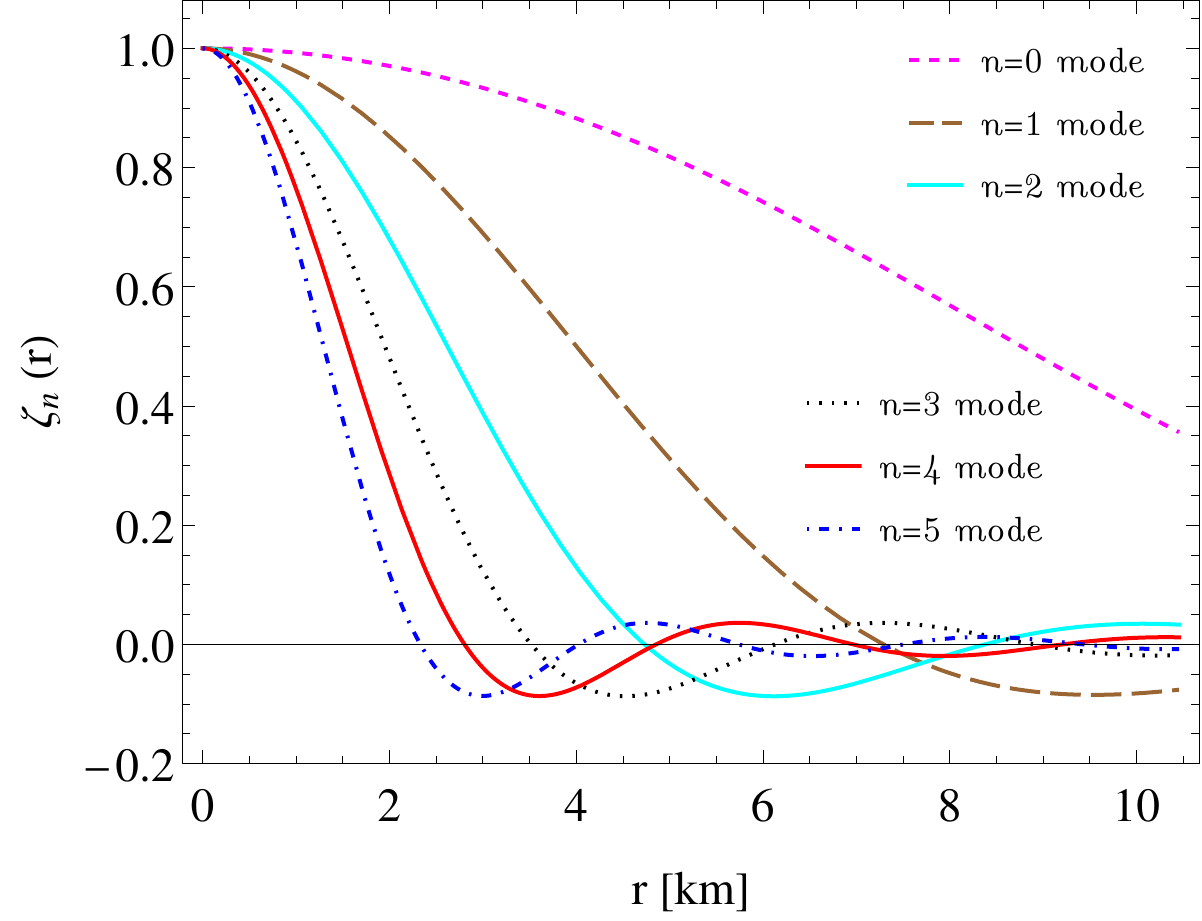}
\includegraphics[width=8.66cm]{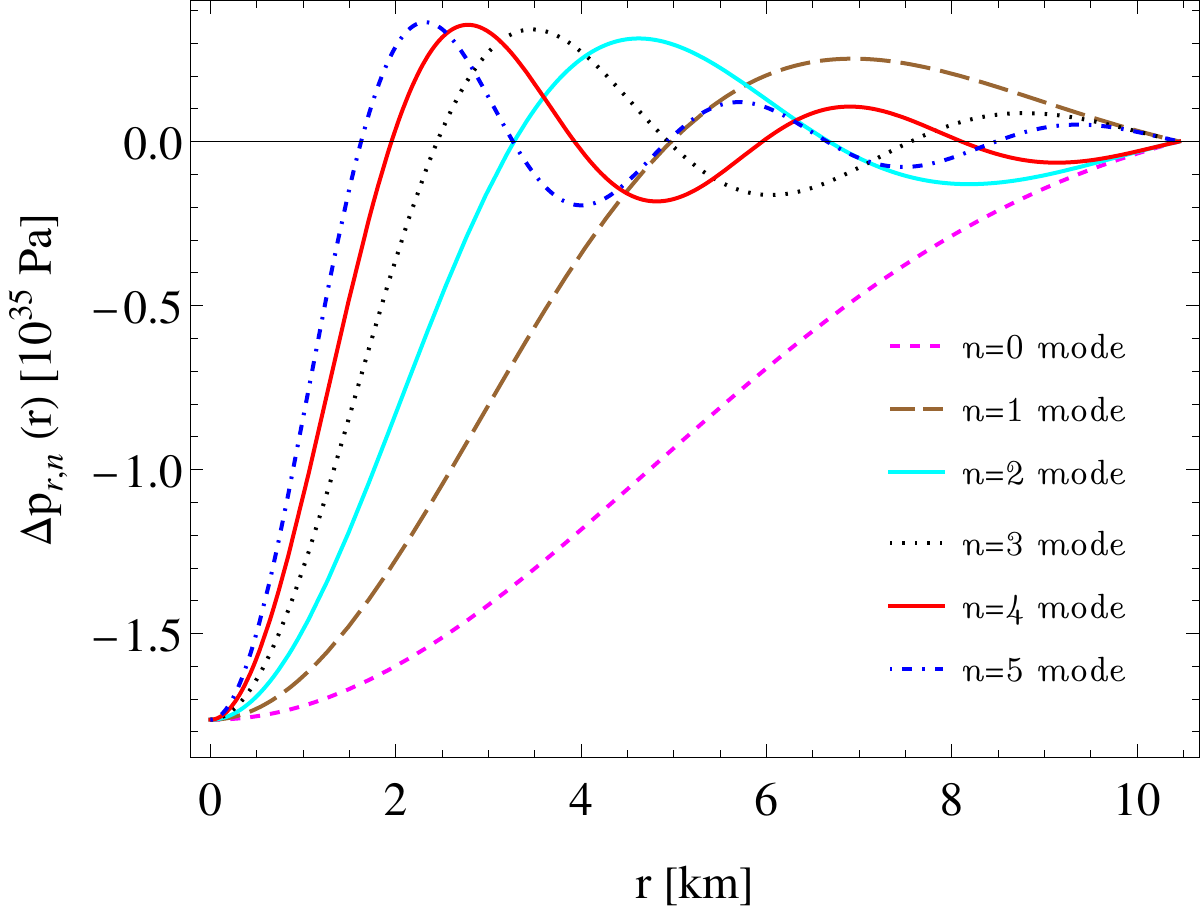}
  \caption{Numerical solution of the radial pulsation equations (\ref{ROEq1}) and (\ref{ROEq2}) in the case of an anisotropic dark energy star with central density $\rho_c = 1.5 \times 10^{18}\, \rm kg/m^3$, $\alpha =0.4$ and EoS parameters given by model I. The radius, mass and the fundamental mode frequency for such configuration are found in Table \ref{table2}. The lines with different colors and styles indicate different overtones so that the solution corresponding to the $n$th vibration mode contains $n$ nodes in the internal structure of the star. Note that the eigenfunctions $\zeta_n(r)$ have been normalized assuming $\zeta =1$ at $r= 0$, and the Lagrangian perturbation of the radial pressure $\Delta p_{r,n}(r)$ obeys the boundary condition (\ref{BCRO2}) at the stellar surface. Since $f_0$ is real, this configuration corresponds to a stable anisotropic dark energy star. }
\label{figure5}
\end{figure*}

Once the equilibrium equations (\ref{TOV1})-(\ref{TOV3}) are integrated from the center to the surface of the star, we then proceed to solve the radial pulsation equations (\ref{ROEq1}) and (\ref{ROEq2}) with the corresponding boundary conditions (\ref{BCRO1}) and (\ref{BCRO2}) using the shooting method. Namely, we integrate from the origin (where we consider the normalized eigenfunctions $\zeta(0)= 1$) up to the stellar surface for a set of trial values $\nu^2$ satisfying the condition (\ref{BCRO1}). In this way, the appropriate eigenfrequencies correspond to the values for which the boundary condition (\ref{BCRO2}) is fulfilled. For instance, for a central density $\rho_c = 1.5 \times 10^{18}\, \rm kg/m^3$, $\alpha =0.4$ and parameters given by model I, Fig.~\ref{figure5} displays the radial behavior of the perturbation variables for the first five squared eigenfrequencies $\nu_n^2$, where $n$ indicates the number of nodes inside the star. This frequency spectrum forms an infinite discrete sequence, i.e. $\nu_0^2< \nu_1^2< \nu_2^2< \cdots$, where the eigenvalue corresponding to $n=0$ is the lowest one (or equivalently, the longest period of all the allowed vibration modes) and it is known as the fundamental mode. Such mode has no nodes, whereas the ﬁrst overtone ($n=1$) has one node, the second overtone ($n=2$) has two, and so on. Stable stars are described by their oscillatory behavior so that $\nu_n^2 >0$ (i.e., $\nu_n$ is purely real). On the other hand, if any of these is negative for a particular star, the frequency is purely imaginary and hence the star is unstable.

\begin{figure*} 
\includegraphics[width=8.63cm]{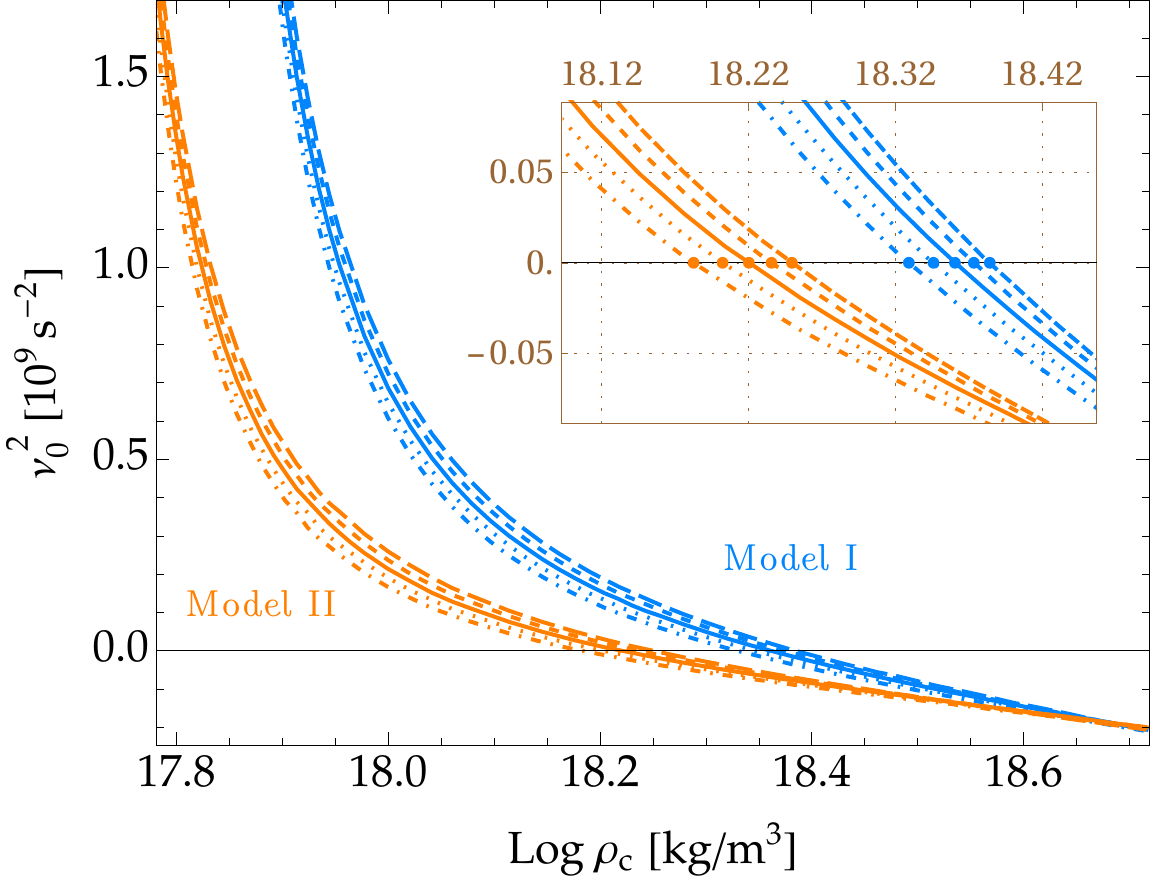}
\includegraphics[width=8.7cm]{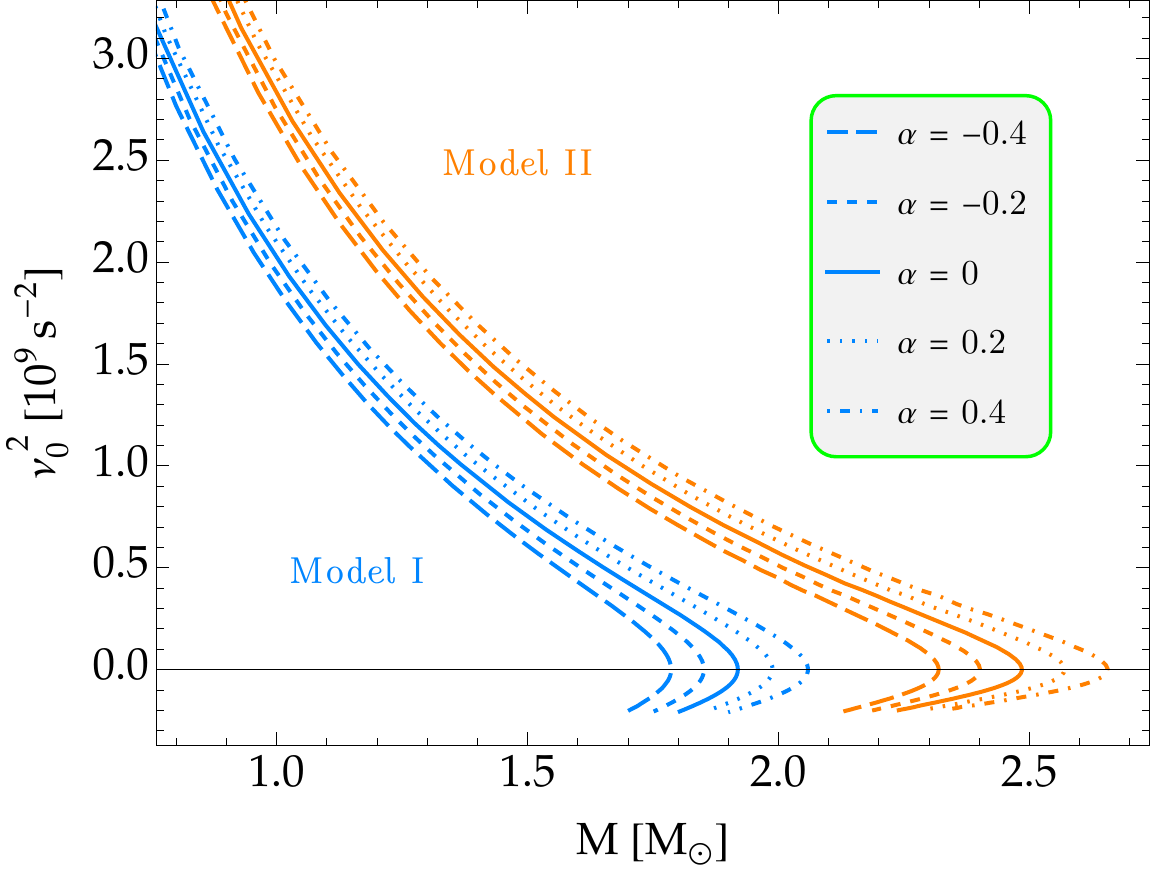}
  \caption{Left panel: Squared frequency of the fundamental pulsation mode as a function of central mass density for anisotropic dark energy stars predicted by Einstein gravity. The full blue and orange circles indicate the central density values where $\nu_0^2 =0$, whose values precisely correspond to the maximum-mass points on the $M(\rho_c)$ curves on the right plot of Fig.~\ref{figure3}. Right plot: Squared frequency of the fundamental mode versus gravitational mass, where it can be observed that the maximum-mass values determine the boundary between stable and unstable stars. }
\label{figure6}
\end{figure*}

Since each higher-order mode has a squared eigenfrequency that is larger than in the case of the preceding mode, it is enough to calculate the frequency of the fundamental pulsation mode for the equilibrium sequences presented in Fig.~\ref{figure3}. With this in mind, in Fig.~\ref{figure6} we plot the squared frequency of the fundamental oscillation mode as a function of the central density (left panel) and gravitational mass (right panel). According to the left plot, the squared frequency of the fundamental mode is exactly zero at the critical-central-density value corresponding to the maximum-mass configuration as shown in the right plot of Fig.~\ref{figure3}, see the full blue and orange circles for both models. Furthermore, according to the right plot of Fig.~\ref{figure6}, the maximum-mass values (that is, when $dM/d\rho_c =0$) can be used as turning points from stability to dynamical instability. Therefore, we can conclude that the usual criterion to guarantee stability $dM/d\rho_c >0$ is still valid for the case of anisotropic dark energy stars. In other words, the conventional $M(\rho_c)$ method is compatible with the calculation of the eigenfrequencies of the normal vibration modes.

If the anisotropic dark energy star has a central density higher than one corresponding to the maximum-mass configuration (indicated by full blue and orange circles in Figs.~\ref{figure3} and \ref{figure6}), the star will become unstable against radial perturbations and collapse to form a black hole. For further details on the dissipative gravitational collapse of compact stellar objects we also refer the reader to Refs.~\cite{Pretel2020EPJC, Pretel2020MNRAS, Bogadi2021, Bogadi2022}. Nonetheless, we must point out that there are EoS models that allow a compact star to migrate to another branch of stable solutions instead of forming a black hole when it is subjected to a perturbation. As a matter of fact, the first-order phase transition between nuclear and quark matter can generate multiple stable branches in the mass-radius diagram for hybrid stars \cite{Alford2013}.

\subsection{Moment of inertia }

\begin{figure*} 
\includegraphics[width=8.3cm]{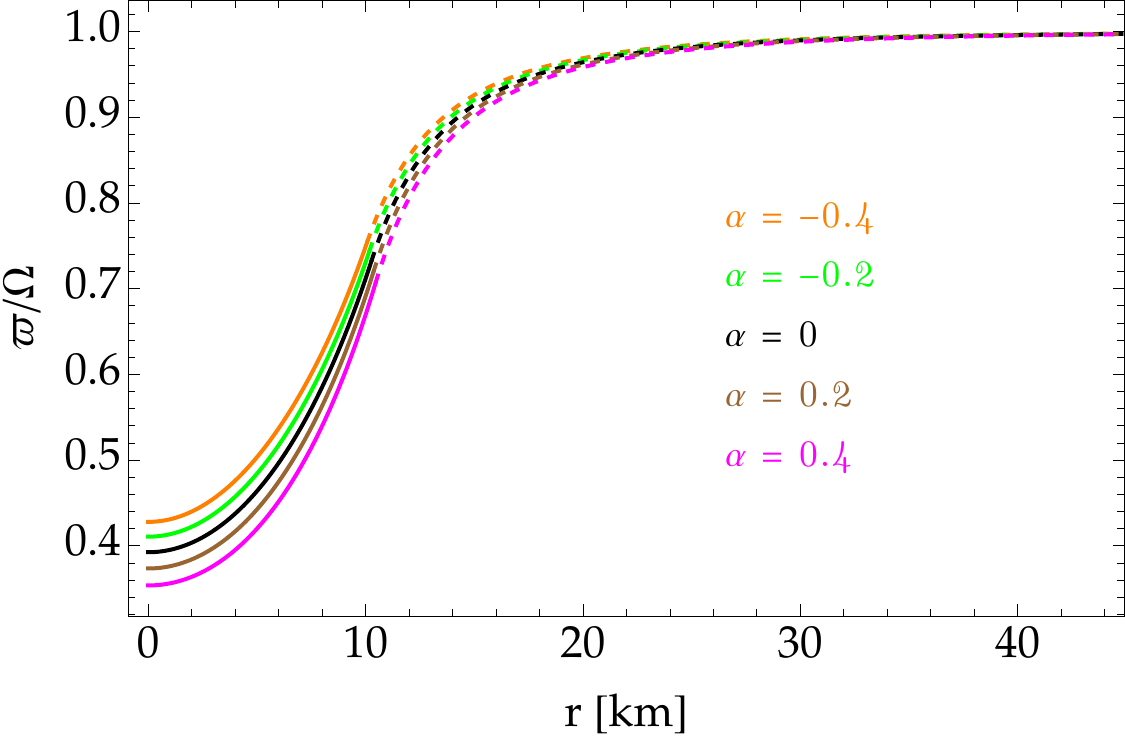}
\includegraphics[width=8.3cm]{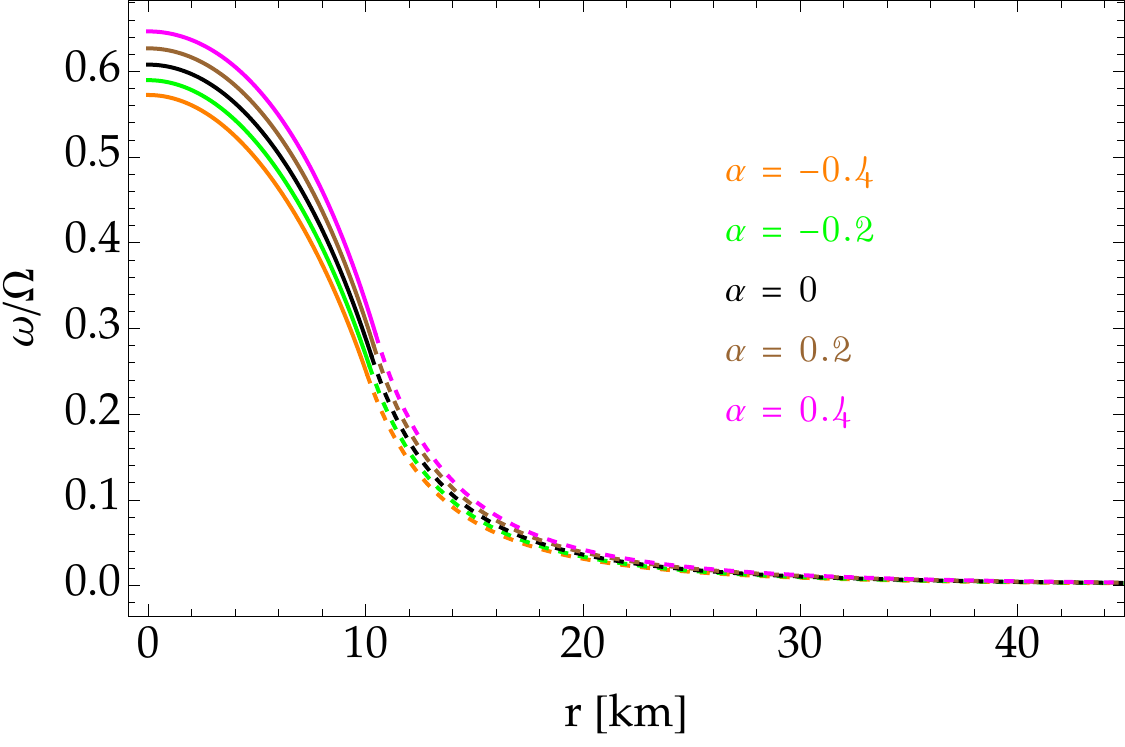}
  \caption{Left panel: Numerical solution of the differential equation (\ref{OmegaEq2}) for a dark energy star described by model I and central density $\rho_c = 1.5 \times 10^{18}\, \rm kg/m^3$ in the presence of anisotropy for several values of the free parameter $\alpha$. The solid and dashed lines represent the interior and exterior solutions, respectively. Right panel: Ratio of frame-dragging angular velocity to the angular velocity of the star, namely $\omega(r)/\Omega = 1- \varpi(r)/\Omega$. It can be observed that the outer solution behaves asymptotically at large distances from the surface of the star (this is, $\omega \rightarrow 0$ for $r \rightarrow \infty$). Furthermore, appreciable changes in the angular velocity due to anisotropy can be noticeable, mainly in the interior region of the star. }
\label{figure7}
\end{figure*}

To calculate the moment of inertia of anisotropic dark energy stars, we first need to solve the differential equation for the rotational drag (\ref{OmegaEq2}) with boundary conditions (\ref{BCMomIner}). In particular, for model I and central density $\rho_c = 1.5 \times 10^{18}\, \rm kg/m^3$, figure \ref{figure7} illustrates the angular velocity everywhere for several values of $\alpha$. As can be observed in the right plot, the dragging angular velocity outside the star has the behavior $\omega(r) \sim r^{-3}$, so that at infinity (where spacetime is flat) the distant local inertial frames do not rotate around the star, namely, $\omega(r) \rightarrow 0$ for $r \rightarrow \infty$. Moreover, anisotropy significantly affects the angular velocity of the local inertial frames in the interior region of the star. More specifically, the dragging angular velocity increases (decreases) for positive (negative) values of the anisotropy parameter $\alpha$. We can then determine the moment of inertia using the integral given in Eq.~(\ref{MomInerEq}). For the above central density, we present the moment of inertia of some dark energy configurations for both models in Table \ref{table2}, where it can be noticed that $I$ increases as the value of $\alpha$ increases.

\begin{figure*} 
\includegraphics[width=8.17cm]{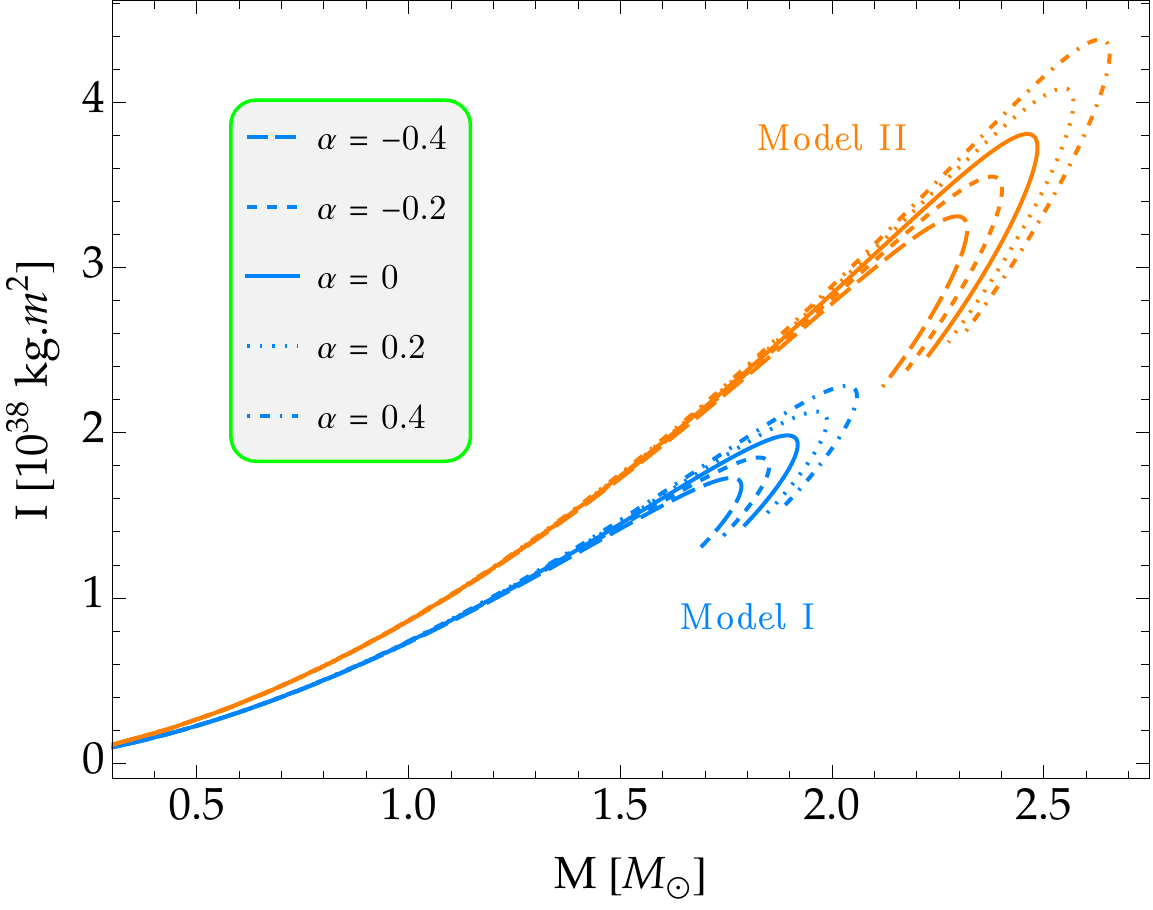}
\includegraphics[width=8.6cm]{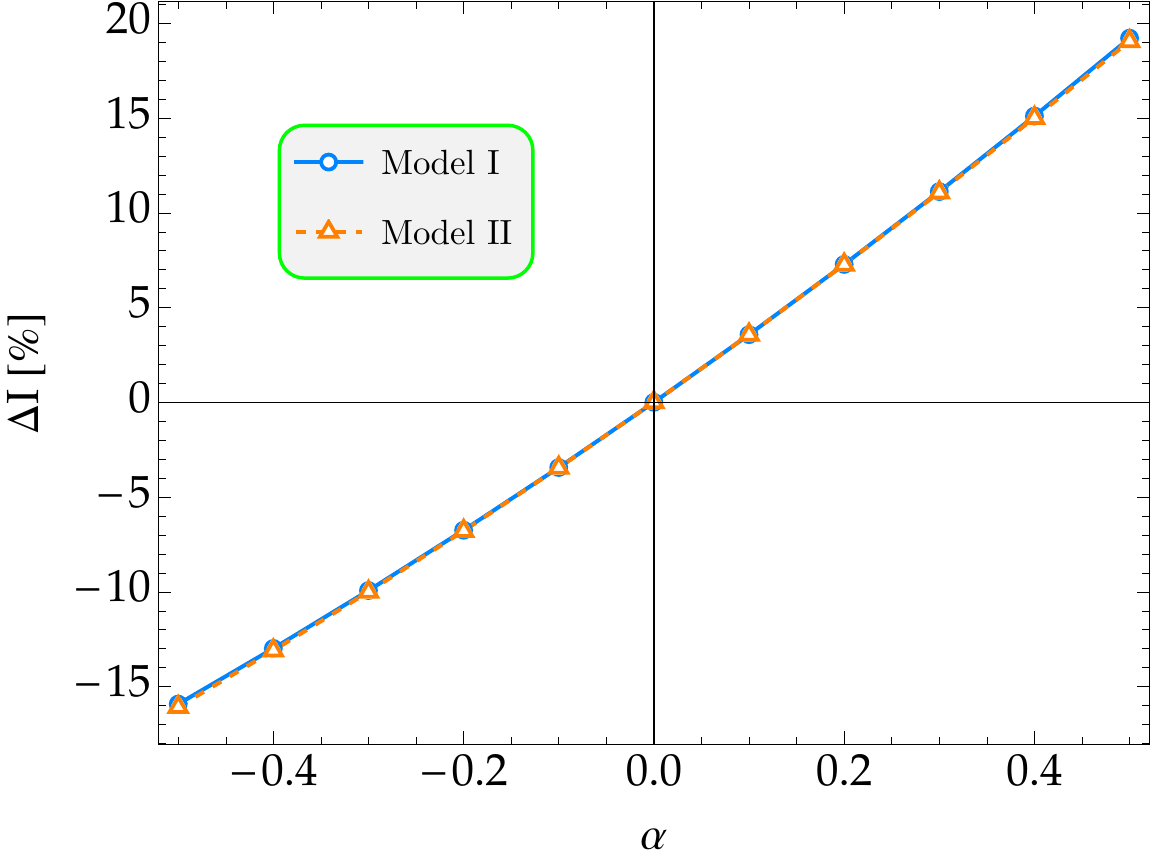}
  \caption{Left panel: Moment of inertia versus mass for anisotropic dark energy stars, where a higher mass results in larger moment on inertia for both models. It is observed that the substantial impact of anisotropy on the moment of inertia occurs predominantly in the high-mass branch. Right panel: Relative deviation (\ref{DeltaIEq}) as a function of the anisotropy parameter. The maximum value of the moment of inertia can undergo variations with respect to its isotropic counterpart of up to $\sim 20\%$ for $\alpha= 0.5$. }
\label{figure8}
\end{figure*}

We can now calculate the moment of inertia for a whole sequence of dark energy stars by varying the central density $\rho_c$. The left panel of Fig.~\ref{figure8} displays the moment of inertia as a function of the gravitational mass for both models. Remarkably, model II provides larger values for the moment of inertia than model I. Indeed, the maximum value $I_{\rm max}$ depends quite sensitively on the free parameters $A$ and $B$ in the EoS (\ref{EoS}). In addition, the main effect of anisotropy on the moment of inertia for slow rotation occurs in the high-mass region, while its influence is irrelevant for sufficiently low masses. In order to better quantify the changes in the maximum values of the moment of inertia induced by the anisotropic pressure, we can define the following relative difference
\begin{equation}\label{DeltaIEq}
    \Delta I = \frac{I_{\rm max,ani} - I_{\rm max,iso}}{I_{\rm max,iso}} ,
\end{equation}
where $I_{\rm max,iso}$ and $I_{\rm max,ani}$ are the maximum values of the moment of inertia for isotropic and anisotropic configurations, respectively. In the right plot of Fig.~\ref{figure8} we present the dependence $\Delta I$ against the anisotropy parameter $\alpha$. The impact of anisotropy is getting stronger as $\vert\alpha\vert$ grows, reaching variations (with respect to the isotropic case) of up to $\sim 20\%$ for $\alpha= 0.5$. We can also note that such relative variations are almost independent of the model adopted.

\subsection{Tidal properties}

\begin{figure*} 
\includegraphics[width=8.405cm]{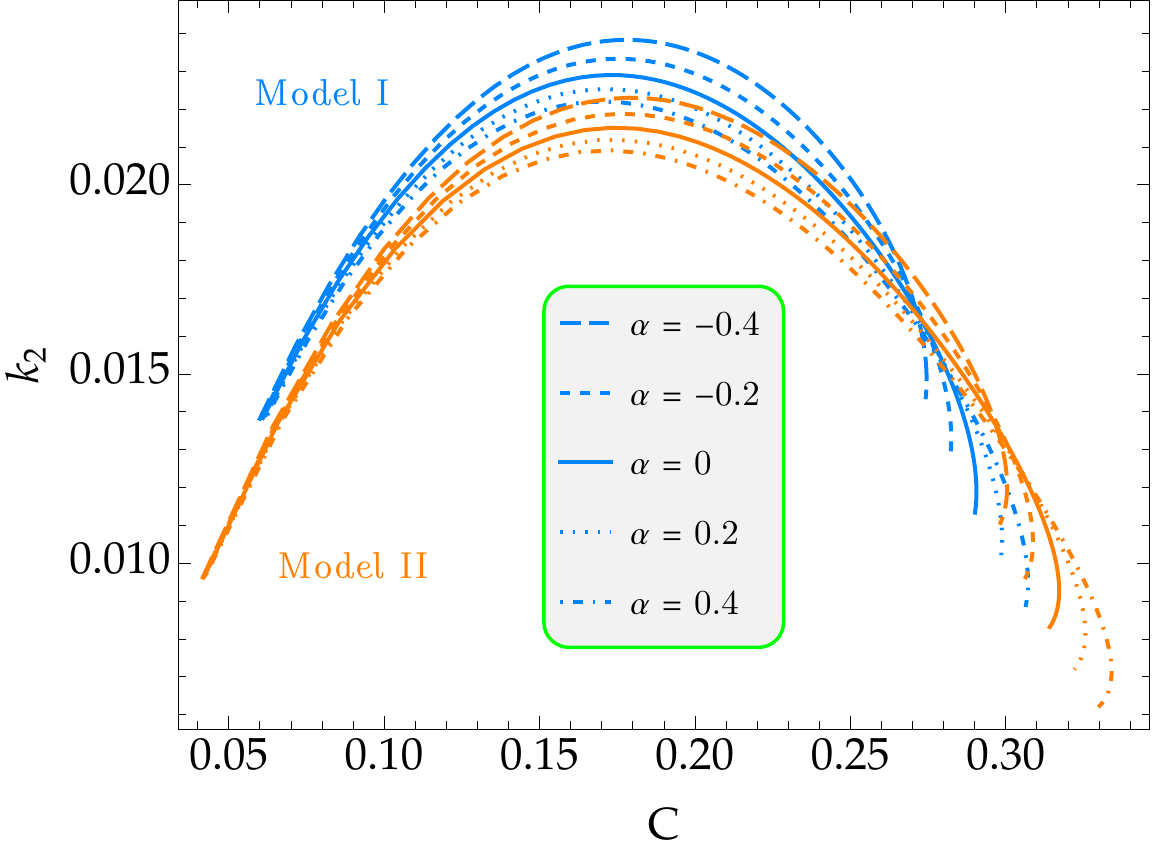}
\includegraphics[width=8.3cm]{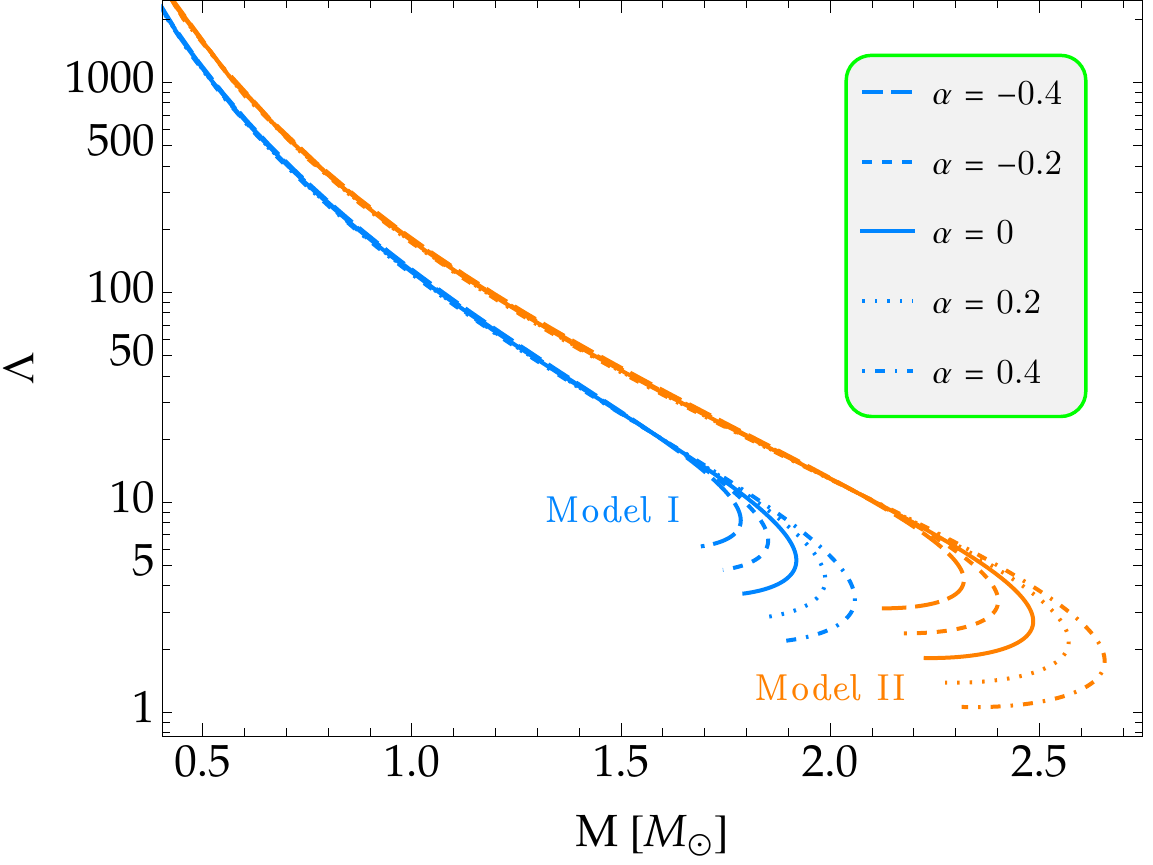}
  \caption{Left panel: Tidal Love number plotted as a function of the compactness $C \equiv M/R$. Right panel: Dimensionless tidal deformability versus gravitational mass predicted by each model, where larger masses yield smaller deformabilities. Note also that the Love number is substantially modified by the anisotropy parameter $\alpha$ for both models, while its greatest effect on tidal deformability $\Lambda$ occurs only in the high-mass region. }
\label{figure9}
\end{figure*}

We will now investigate how the anisotropy parameter $\alpha$ affects the tidal properties of dark energy stars. Given a specific value of $\alpha$, this requires solving the differential equation (\ref{yEq}) for a range of central densities. The left panel of Fig.~\ref{figure9} is the result of calculating the tidal Love number (\ref{LoveNumEq}) for a sequence of stellar configurations by considering different values of $\alpha$, where the isotropic case corresponds to $\alpha =0$. Similar to the trends in strange quark stars, as reported in Ref.~\cite{Kumari2021}, the Love number of dark energy stars grows until it reaches a maximum value and then decreases as compactness increases. Note also that the maximum value of $k_2$ is sensitive to the value of $\alpha$, indicating that the Love number decreases as the parameter $\alpha$ increases for both models. Although model II provides larger maximum masses (as well as redshift and moment of inertia) than model I, we see that the behavior is different for the maximum values in the tidal Love number.

Ultimately, in the right plot of Fig.~\ref{figure9}, the dimensionless tidal deformability $\Lambda= \bar{\lambda}/M^5$ is plotted as a function of mass, where it can be observed that smaller masses yield higher deformabilities. In each model, the presence of anisotropy has a negligible effect on $\Lambda$ for small masses, while slightly more significant changes take place only in the high-mass region.

\section{Conclusions and outlook }\label{Sec5}

In this work, we have focused on the equilibrium structure of dark energy stars by using a Chaplygin-like equation of state under the presence of both isotropic and anisotropic pressures within the context of standard GR. Our goal was to construct stable compact stars whose characteristics could be compared with the observational data on the mass-radius diagram. In this perspective, the global properties of a compact star such as radius, mass, redshift, moment of inertia, oscillation spectrum and tidal deformability have been calculated. To describe the anisotropic pressure within the dark energy fluid we have adopted the anisotropy profile proposed by Horvat \textit{et al.}~\cite{Horvat2011}, where a free parameter $\alpha$ measures the degree of anisotropy.

We have discussed the possibility of observing stable dark energy stars made of a negative pressure fluid ``$-B/\rho$'' plus a barotropic component ``$A\rho$''. By way of comparison, the EoS parameters $A$ and $B$ have been chosen in such a way that they agree sufficiently with the observational data, e.g.~the mass-radius constraint from the GW170817 event. For isotropic configurations, we have shown that various sets of values $\lbrace A, B\rbrace$ can be chosen since they obey the causality condition and consistently describe compact stars observed in the Universe. Furthermore, we saw that the secondary component resulting from the gravitational-wave signal GW190814 \citep{Abbott2020} can be described as a dark energy star using $A = 0.4$ and $B \in [4, 5]\mu$.

Based on these results, we have established two models with different values $A$ and $B$ in order to explore the effects of anisotropy in the interior region of a dark energy star. In particular, the maximum-mass values increase as the parameter $\alpha$ increases. We noticed that model I without anisotropic pressures is not capable of generating maximum masses above $2M_\odot$. However, the inclusion of anisotropies ($\alpha = 0.4$) allows a significant increase in the maximum mass and thus a more favorable description of the compact objects observed in nature. On the other hand, model II with anisotropies fits better with the observational measurements, although such a model can lead to the formation of ultra-compact objects for sufficiently large values of $\alpha$. We also calculated the surface gravitational redshift for such stars, and our results indicated that $z_{\rm sur}$ is substantially affected by the anisotropy in the high-mass branch, while the changes are irrelevant for sufficiently low masses.

A star exists in the Universe only if it is dynamically stable, so our second task was to investigate whether the dark energy stars are stable or unstable with respect to an adiabatic radial perturbation. Our results showed that the standard criterion for radial stability $dM/d\rho_c >0$ still holds for dark energy stars since the squared frequency of the fundamental pulsation mode ($\nu_0^2$) vanishes at the critical central density corresponding to the maximum-mass configuration. This has been examined in detail for both isotropic ($\alpha =0$) and anisotropic ($\alpha \neq 0$) stellar configurations.

In the slowly rotating approximation, where only first-order terms in the angular velocity are kept, we have also determined the moment of inertia of anisotropic dark energy stars. For this purpose, we first had to calculate the frame-dragging angular velocity for each central density. The presence of anisotropic pressure results in a substantial increase (decrease) of the angular velocity $\omega$ for more positive (negative) values of $\alpha$. We found that the significant impact of the anisotropy on the moment of inertia occurs mainly in the high-mass branch for both models. Furthermore, the maximum value of the moment of inertia can undergo variations of up to $\sim 20\%$ for $\alpha= 0.5$ as compared with the isotropic case.

We have analyzed the effect of anisotropic pressure on the tidal properties of such stars. In particular, our outcomes revealed that the tidal Love number is sensitive to moderate variations of the parameter $\alpha$, indicating that the maximum value of $k_2$ can increase as $\alpha$ decreases. In addition, the greatest effect of anisotropy on the dimensionless tidal deformability takes place only in the high-mass region. Based on the foregoing results, the present work thereby serves to develop a comprehensive perspective on the relativistic structure of dark energy stars in the presence of anisotropy.

Summarizing, we have explored the possible existence of stable dark energy stars whose masses and radii are not in disagreement with the current observational data. The Chaplygin-like EoS predicts maximum-mass values consistent with observational measurements of highly massive pulsars. Future research includes the adoption of widespread versions of Chaplygin gas that best fit key cosmological parameters. In future studies we will thereby take further steps in that direction, focusing on the different types of generalized Chaplygin gas models as discussed in Ref.~\cite{Zheng2022}. In addition, as carried out in the case of boson stars \cite{Sennett2017}, it would be interesting to employ a Fisher matrix analysis in order to distinguish dark energy stars from black holes and neutron stars from tidal interactions in inspiraling binary systems. It is also worth mentioning that Romano \cite{Romano2022} has recently discussed the effects of dark energy on the propagation of gravitational waves. In that regard, we expect that future electromagnetic observations of compact binaries and gravitational-wave astronomy will provide a better understanding of compact stars in the presence of dark energy, and even help us answer the most basic question: How did dark energy form in the Universe? Anyway, our results suggest that dark energy stars deserve further investigation by taking into account the cosmological aspects as well as the gravitational-wave signals from binary mergers.

\begin{acknowledgments}
The author would like to acknowledge the anonymous reviewer for useful constructive feedback and valuable suggestions. The author would also like to thank Maria F. A. da Silva for giving helpful comments. This research work was financially supported by the PCI program of the Brazilian agency ``Conselho Nacional de Desenvolvimento Científico e Tecnológico''--CNPq. 
\end{acknowledgments}\


%


\newpage
\begin{thebibliography}{102}%
\makeatletter
\providecommand \@ifxundefined [1]{%
 \@ifx{#1\undefined}
}%
\providecommand \@ifnum [1]{%
 \ifnum #1\expandafter \@firstoftwo
 \else \expandafter \@secondoftwo
 \fi
}%
\providecommand \@ifx [1]{%
 \ifx #1\expandafter \@firstoftwo
 \else \expandafter \@secondoftwo
 \fi
}%
\providecommand \natexlab [1]{#1}%
\providecommand \enquote  [1]{``#1''}%
\providecommand \bibnamefont  [1]{#1}%
\providecommand \bibfnamefont [1]{#1}%
\providecommand \citenamefont [1]{#1}%
\providecommand \href@noop [0]{\@secondoftwo}%
\providecommand \href [0]{\begingroup \@sanitize@url \@href}%
\providecommand \@href[1]{\@@startlink{#1}\@@href}%
\providecommand \@@href[1]{\endgroup#1\@@endlink}%
\providecommand \@sanitize@url [0]{\catcode `\\12\catcode `\$12\catcode
  `\&12\catcode `\#12\catcode `\^12\catcode `\_12\catcode `\%12\relax}%
\providecommand \@@startlink[1]{}%
\providecommand \@@endlink[0]{}%
\providecommand \url  [0]{\begingroup\@sanitize@url \@url }%
\providecommand \@url [1]{\endgroup\@href {#1}{\urlprefix }}%
\providecommand \urlprefix  [0]{URL }%
\providecommand \Eprint [0]{\href }%
\providecommand \doibase [0]{http://dx.doi.org/}%
\providecommand \selectlanguage [0]{\@gobble}%
\providecommand \bibinfo  [0]{\@secondoftwo}%
\providecommand \bibfield  [0]{\@secondoftwo}%
\providecommand \translation [1]{[#1]}%
\providecommand \BibitemOpen [0]{}%
\providecommand \bibitemStop [0]{}%
\providecommand \bibitemNoStop [0]{.\EOS\space}%
\providecommand \EOS [0]{\spacefactor3000\relax}%
\providecommand \BibitemShut  [1]{\csname bibitem#1\endcsname}%
\let\auto@bib@innerbib\@empty
\bibitem [{\citenamefont {Aghanim}\ \emph {et~al.}(2020)\citenamefont {Aghanim}
  \emph {et~al.}}]{Aghanim2020}%
  \BibitemOpen
  \bibfield  {author} {\bibinfo {author} {\bibfnamefont {N.}~\bibnamefont
  {Aghanim}} \emph {et~al.},\ }\href {\doibase 10.1051/0004-6361/201833910}
  {\bibfield  {journal} {\bibinfo  {journal} {A\&A}\ }\textbf {\bibinfo
  {volume} {641}},\ \bibinfo {pages} {A6} (\bibinfo {year} {2020})}\BibitemShut
  {NoStop}%
\bibitem [{\citenamefont {Weinberg}(1989)}]{Weinberg1989}%
  \BibitemOpen
  \bibfield  {author} {\bibinfo {author} {\bibfnamefont {S.}~\bibnamefont
  {Weinberg}},\ }\href {\doibase 10.1103/RevModPhys.61.1} {\bibfield  {journal}
  {\bibinfo  {journal} {Rev. Mod. Phys.}\ }\textbf {\bibinfo {volume} {61}},\
  \bibinfo {pages} {1} (\bibinfo {year} {1989})}\BibitemShut {NoStop}%
\bibitem [{\citenamefont {Padmanabhan}(2003)}]{Padmanabhan2003}%
  \BibitemOpen
  \bibfield  {author} {\bibinfo {author} {\bibfnamefont {T.}~\bibnamefont
  {Padmanabhan}},\ }\href {\doibase
  https://doi.org/10.1016/S0370-1573(03)00120-0} {\bibfield  {journal}
  {\bibinfo  {journal} {Physics Reports}\ }\textbf {\bibinfo {volume} {380}},\
  \bibinfo {pages} {235} (\bibinfo {year} {2003})}\BibitemShut {NoStop}%
\bibitem [{\citenamefont {Kamenshchik}\ \emph {et~al.}(2001)\citenamefont
  {Kamenshchik}, \citenamefont {Moschella},\ and\ \citenamefont
  {Pasquier}}]{KAMENSHCHIK2001265}%
  \BibitemOpen
  \bibfield  {author} {\bibinfo {author} {\bibfnamefont {A.}~\bibnamefont
  {Kamenshchik}}, \bibinfo {author} {\bibfnamefont {U.}~\bibnamefont
  {Moschella}}, \ and\ \bibinfo {author} {\bibfnamefont {V.}~\bibnamefont
  {Pasquier}},\ }\href {\doibase https://doi.org/10.1016/S0370-2693(01)00571-8}
  {\bibfield  {journal} {\bibinfo  {journal} {Physics Letters B}\ }\textbf
  {\bibinfo {volume} {511}},\ \bibinfo {pages} {265} (\bibinfo {year}
  {2001})}\BibitemShut {NoStop}%
\bibitem [{\citenamefont {Bento}\ \emph {et~al.}(2002)\citenamefont {Bento},
  \citenamefont {Bertolami},\ and\ \citenamefont {Sen}}]{Bento2002}%
  \BibitemOpen
  \bibfield  {author} {\bibinfo {author} {\bibfnamefont {M.~C.}\ \bibnamefont
  {Bento}}, \bibinfo {author} {\bibfnamefont {O.}~\bibnamefont {Bertolami}}, \
  and\ \bibinfo {author} {\bibfnamefont {A.~A.}\ \bibnamefont {Sen}},\ }\href
  {\doibase 10.1103/PhysRevD.66.043507} {\bibfield  {journal} {\bibinfo
  {journal} {Phys. Rev. D}\ }\textbf {\bibinfo {volume} {66}},\ \bibinfo
  {pages} {043507} (\bibinfo {year} {2002})}\BibitemShut {NoStop}%
\bibitem [{\citenamefont {Reis}\ \emph {et~al.}(2003)\citenamefont {Reis},
  \citenamefont {Waga}, \citenamefont {Calv\~ao},\ and\ \citenamefont
  {Jor\'as}}]{Reis2003}%
  \BibitemOpen
  \bibfield  {author} {\bibinfo {author} {\bibfnamefont {R.~R.~R.}\
  \bibnamefont {Reis}}, \bibinfo {author} {\bibfnamefont {I.}~\bibnamefont
  {Waga}}, \bibinfo {author} {\bibfnamefont {M.~O.}\ \bibnamefont {Calv\~ao}},
  \ and\ \bibinfo {author} {\bibfnamefont {S.~E.}\ \bibnamefont {Jor\'as}},\
  }\href {\doibase 10.1103/PhysRevD.68.061302} {\bibfield  {journal} {\bibinfo
  {journal} {Phys. Rev. D}\ }\textbf {\bibinfo {volume} {68}},\ \bibinfo
  {pages} {061302} (\bibinfo {year} {2003})}\BibitemShut {NoStop}%
\bibitem [{\citenamefont {Xu}\ \emph {et~al.}(2012)\citenamefont {Xu},
  \citenamefont {Lu},\ and\ \citenamefont {Wang}}]{Xu2012}%
  \BibitemOpen
  \bibfield  {author} {\bibinfo {author} {\bibfnamefont {L.}~\bibnamefont
  {Xu}}, \bibinfo {author} {\bibfnamefont {J.}~\bibnamefont {Lu}}, \ and\
  \bibinfo {author} {\bibfnamefont {Y.}~\bibnamefont {Wang}},\ }\href {\doibase
  https://doi.org/10.1140/epjc/s10052-012-1883-7} {\bibfield  {journal}
  {\bibinfo  {journal} {Eur. Phys. J. C}\ }\textbf {\bibinfo {volume} {72}},\
  \bibinfo {pages} {1883} (\bibinfo {year} {2012})}\BibitemShut {NoStop}%
\bibitem [{\citenamefont {Bilić}\ \emph {et~al.}(2002)\citenamefont {Bilić},
  \citenamefont {Tupper},\ and\ \citenamefont {Viollier}}]{BILIC2002}%
  \BibitemOpen
  \bibfield  {author} {\bibinfo {author} {\bibfnamefont {N.}~\bibnamefont
  {Bilić}}, \bibinfo {author} {\bibfnamefont {G.}~\bibnamefont {Tupper}}, \
  and\ \bibinfo {author} {\bibfnamefont {R.}~\bibnamefont {Viollier}},\ }\href
  {\doibase https://doi.org/10.1016/S0370-2693(02)01716-1} {\bibfield
  {journal} {\bibinfo  {journal} {Physics Letters B}\ }\textbf {\bibinfo
  {volume} {535}},\ \bibinfo {pages} {17} (\bibinfo {year} {2002})}\BibitemShut
  {NoStop}%
\bibitem [{\citenamefont {Makler}\ \emph {et~al.}(2003)\citenamefont {Makler},
  \citenamefont {{de Oliveira}},\ and\ \citenamefont {Waga}}]{Makler2003}%
  \BibitemOpen
  \bibfield  {author} {\bibinfo {author} {\bibfnamefont {M.}~\bibnamefont
  {Makler}}, \bibinfo {author} {\bibfnamefont {S.~Q.}\ \bibnamefont {{de
  Oliveira}}}, \ and\ \bibinfo {author} {\bibfnamefont {I.}~\bibnamefont
  {Waga}},\ }\href {\doibase https://doi.org/10.1016/S0370-2693(03)00038-8}
  {\bibfield  {journal} {\bibinfo  {journal} {Physics Letters B}\ }\textbf
  {\bibinfo {volume} {555}},\ \bibinfo {pages} {1} (\bibinfo {year}
  {2003})}\BibitemShut {NoStop}%
\bibitem [{\citenamefont {Li}\ \emph {et~al.}(2019)\citenamefont {Li},
  \citenamefont {Yang},\ and\ \citenamefont {Gai}}]{Li2019}%
  \BibitemOpen
  \bibfield  {author} {\bibinfo {author} {\bibfnamefont {H.}~\bibnamefont
  {Li}}, \bibinfo {author} {\bibfnamefont {W.}~\bibnamefont {Yang}}, \ and\
  \bibinfo {author} {\bibfnamefont {L.}~\bibnamefont {Gai}},\ }\href {\doibase
  https://doi.org/10.1051/0004-6361/201833836} {\bibfield  {journal} {\bibinfo
  {journal} {A\&A}\ }\textbf {\bibinfo {volume} {623}},\ \bibinfo {pages} {A28}
  (\bibinfo {year} {2019})}\BibitemShut {NoStop}%
\bibitem [{\citenamefont {Zheng}\ \emph {et~al.}(2022)\citenamefont {Zheng}
  \emph {et~al.}}]{Zheng2022}%
  \BibitemOpen
  \bibfield  {author} {\bibinfo {author} {\bibfnamefont {J.}~\bibnamefont
  {Zheng}} \emph {et~al.},\ }\href {\doibase
  https://doi.org/10.1140/epjc/s10052-022-10517-4} {\bibfield  {journal}
  {\bibinfo  {journal} {Eur. Phys. J. C}\ }\textbf {\bibinfo {volume} {82}},\
  \bibinfo {pages} {582} (\bibinfo {year} {2022})}\BibitemShut {NoStop}%
\bibitem [{\citenamefont {Ignatov}\ and\ \citenamefont
  {Pieroni}(2021)}]{Ignatov2021}%
  \BibitemOpen
  \bibfield  {author} {\bibinfo {author} {\bibfnamefont {Y.}~\bibnamefont
  {Ignatov}}\ and\ \bibinfo {author} {\bibfnamefont {M.}~\bibnamefont
  {Pieroni}},\ }\href {https://arxiv.org/abs/2110.10085} {\bibfield  {journal}
  {\bibinfo  {journal} {arXiv:2110.10085 [astro-ph.CO]}\ } (\bibinfo {year}
  {2021})}\BibitemShut {NoStop}%
\bibitem [{\citenamefont {Odintsov}\ \emph {et~al.}(2020)\citenamefont
  {Odintsov}, \citenamefont {G\'omez},\ and\ \citenamefont
  {Sharov}}]{OdintsovGS2020}%
  \BibitemOpen
  \bibfield  {author} {\bibinfo {author} {\bibfnamefont {S.~D.}\ \bibnamefont
  {Odintsov}}, \bibinfo {author} {\bibfnamefont {D.~S.-C.}\ \bibnamefont
  {G\'omez}}, \ and\ \bibinfo {author} {\bibfnamefont {G.~S.}\ \bibnamefont
  {Sharov}},\ }\href {\doibase 10.1103/PhysRevD.101.044010} {\bibfield
  {journal} {\bibinfo  {journal} {Phys. Rev. D}\ }\textbf {\bibinfo {volume}
  {101}},\ \bibinfo {pages} {044010} (\bibinfo {year} {2020})}\BibitemShut
  {NoStop}%
\bibitem [{\citenamefont {Copeland}\ \emph {et~al.}(2006)\citenamefont
  {Copeland}, \citenamefont {Sami},\ and\ \citenamefont
  {Tsujikawa}}]{Copeland2006}%
  \BibitemOpen
  \bibfield  {author} {\bibinfo {author} {\bibfnamefont {E.~J.}\ \bibnamefont
  {Copeland}}, \bibinfo {author} {\bibfnamefont {M.}~\bibnamefont {Sami}}, \
  and\ \bibinfo {author} {\bibfnamefont {S.}~\bibnamefont {Tsujikawa}},\ }\href
  {\doibase https://doi.org/10.1142/S021827180600942X} {\bibfield  {journal}
  {\bibinfo  {journal} {Int. J. Mod. Phys. D}\ }\textbf {\bibinfo {volume}
  {15}},\ \bibinfo {pages} {1753} (\bibinfo {year} {2006})}\BibitemShut
  {NoStop}%
\bibitem [{\citenamefont {Koyama}(2016)}]{Koyama2016}%
  \BibitemOpen
  \bibfield  {author} {\bibinfo {author} {\bibfnamefont {K.}~\bibnamefont
  {Koyama}},\ }\href {\doibase 10.1088/0034-4885/79/4/046902} {\bibfield
  {journal} {\bibinfo  {journal} {Rep. Prog. Phys.}\ }\textbf {\bibinfo
  {volume} {79}},\ \bibinfo {pages} {046902} (\bibinfo {year}
  {2016})}\BibitemShut {NoStop}%
\bibitem [{\citenamefont {Boisseau}\ \emph {et~al.}(2000)\citenamefont
  {Boisseau}, \citenamefont {Esposito-Far\`ese}, \citenamefont {Polarski},\
  and\ \citenamefont {Starobinsky}}]{Boisseau2000}%
  \BibitemOpen
  \bibfield  {author} {\bibinfo {author} {\bibfnamefont {B.}~\bibnamefont
  {Boisseau}}, \bibinfo {author} {\bibfnamefont {G.}~\bibnamefont
  {Esposito-Far\`ese}}, \bibinfo {author} {\bibfnamefont {D.}~\bibnamefont
  {Polarski}}, \ and\ \bibinfo {author} {\bibfnamefont {A.~A.}\ \bibnamefont
  {Starobinsky}},\ }\href {\doibase 10.1103/PhysRevLett.85.2236} {\bibfield
  {journal} {\bibinfo  {journal} {Phys. Rev. Lett.}\ }\textbf {\bibinfo
  {volume} {85}},\ \bibinfo {pages} {2236} (\bibinfo {year}
  {2000})}\BibitemShut {NoStop}%
\bibitem [{\citenamefont {Esposito-Far\`ese}\ and\ \citenamefont
  {Polarski}(2001)}]{Esposito2001}%
  \BibitemOpen
  \bibfield  {author} {\bibinfo {author} {\bibfnamefont {G.}~\bibnamefont
  {Esposito-Far\`ese}}\ and\ \bibinfo {author} {\bibfnamefont {D.}~\bibnamefont
  {Polarski}},\ }\href {\doibase 10.1103/PhysRevD.63.063504} {\bibfield
  {journal} {\bibinfo  {journal} {Phys. Rev. D}\ }\textbf {\bibinfo {volume}
  {63}},\ \bibinfo {pages} {063504} (\bibinfo {year} {2001})}\BibitemShut
  {NoStop}%
\bibitem [{\citenamefont {Sotiriou}\ and\ \citenamefont
  {Faraoni}(2010)}]{Sotiriou2010}%
  \BibitemOpen
  \bibfield  {author} {\bibinfo {author} {\bibfnamefont {T.~P.}\ \bibnamefont
  {Sotiriou}}\ and\ \bibinfo {author} {\bibfnamefont {V.}~\bibnamefont
  {Faraoni}},\ }\href {\doibase 10.1103/RevModPhys.82.451} {\bibfield
  {journal} {\bibinfo  {journal} {Rev. Mod. Phys.}\ }\textbf {\bibinfo {volume}
  {82}},\ \bibinfo {pages} {451} (\bibinfo {year} {2010})}\BibitemShut
  {NoStop}%
\bibitem [{\citenamefont {{De Felice}}\ and\ \citenamefont
  {{Tsujikawa}}(2010)}]{Felice2010}%
  \BibitemOpen
  \bibfield  {author} {\bibinfo {author} {\bibfnamefont {A.}~\bibnamefont {{De
  Felice}}}\ and\ \bibinfo {author} {\bibfnamefont {S.}~\bibnamefont
  {{Tsujikawa}}},\ }\href {\doibase 10.12942/lrr-2010-3} {\bibfield  {journal}
  {\bibinfo  {journal} {Living Rev. Relativ.}\ }\textbf {\bibinfo {volume}
  {13}},\ \bibinfo {eid} {3} (\bibinfo {year} {2010})}\BibitemShut {NoStop}%
\bibitem [{\citenamefont {Starobinsky}(1980)}]{Starobinsky1980}%
  \BibitemOpen
  \bibfield  {author} {\bibinfo {author} {\bibfnamefont {A.}~\bibnamefont
  {Starobinsky}},\ }\href {\doibase
  https://doi.org/10.1016/0370-2693(80)90670-X} {\bibfield  {journal} {\bibinfo
   {journal} {Physics Letters B}\ }\textbf {\bibinfo {volume} {91}},\ \bibinfo
  {pages} {99 } (\bibinfo {year} {1980})}\BibitemShut {NoStop}%
\bibitem [{\citenamefont {Carroll}\ \emph {et~al.}(2004)\citenamefont {Carroll}
  \emph {et~al.}}]{Carroll2004}%
  \BibitemOpen
  \bibfield  {author} {\bibinfo {author} {\bibfnamefont {S.~M.}\ \bibnamefont
  {Carroll}} \emph {et~al.},\ }\href {\doibase 10.1103/PhysRevD.70.043528}
  {\bibfield  {journal} {\bibinfo  {journal} {Phys. Rev. D}\ }\textbf {\bibinfo
  {volume} {70}},\ \bibinfo {pages} {043528} (\bibinfo {year}
  {2004})}\BibitemShut {NoStop}%
\bibitem [{\citenamefont {Chiba}(2003)}]{Chiba2003}%
  \BibitemOpen
  \bibfield  {author} {\bibinfo {author} {\bibfnamefont {T.}~\bibnamefont
  {Chiba}},\ }\href {\doibase https://doi.org/10.1016/j.physletb.2003.09.033}
  {\bibfield  {journal} {\bibinfo  {journal} {Physics Letters B}\ }\textbf
  {\bibinfo {volume} {575}},\ \bibinfo {pages} {1} (\bibinfo {year}
  {2003})}\BibitemShut {NoStop}%
\bibitem [{\citenamefont {Nojiri}\ and\ \citenamefont
  {Odintsov}(2011)}]{Nojiri2011}%
  \BibitemOpen
  \bibfield  {author} {\bibinfo {author} {\bibfnamefont {S.}~\bibnamefont
  {Nojiri}}\ and\ \bibinfo {author} {\bibfnamefont {S.~D.}\ \bibnamefont
  {Odintsov}},\ }\href {\doibase https://doi.org/10.1016/j.physrep.2011.04.001}
  {\bibfield  {journal} {\bibinfo  {journal} {Physics Reports}\ }\textbf
  {\bibinfo {volume} {505}},\ \bibinfo {pages} {59} (\bibinfo {year}
  {2011})}\BibitemShut {NoStop}%
\bibitem [{\citenamefont {Clifton}\ \emph {et~al.}(2012)\citenamefont {Clifton}
  \emph {et~al.}}]{Clifton2012}%
  \BibitemOpen
  \bibfield  {author} {\bibinfo {author} {\bibfnamefont {T.}~\bibnamefont
  {Clifton}} \emph {et~al.},\ }\href {\doibase
  https://doi.org/10.1016/j.physrep.2012.01.001} {\bibfield  {journal}
  {\bibinfo  {journal} {Physics Reports}\ }\textbf {\bibinfo {volume} {513}},\
  \bibinfo {pages} {1} (\bibinfo {year} {2012})}\BibitemShut {NoStop}%
\bibitem [{\citenamefont {Nojiri}\ \emph {et~al.}(2017)\citenamefont {Nojiri},
  \citenamefont {Odintsov},\ and\ \citenamefont {Oikonomou}}]{Nojiri2017}%
  \BibitemOpen
  \bibfield  {author} {\bibinfo {author} {\bibfnamefont {S.}~\bibnamefont
  {Nojiri}}, \bibinfo {author} {\bibfnamefont {S.}~\bibnamefont {Odintsov}}, \
  and\ \bibinfo {author} {\bibfnamefont {V.}~\bibnamefont {Oikonomou}},\ }\href
  {\doibase https://doi.org/10.1016/j.physrep.2017.06.001} {\bibfield
  {journal} {\bibinfo  {journal} {Physics Reports}\ }\textbf {\bibinfo {volume}
  {692}},\ \bibinfo {pages} {1} (\bibinfo {year} {2017})}\BibitemShut {NoStop}%
\bibitem [{\citenamefont {Folomeev}(2018)}]{Folomeev2018}%
  \BibitemOpen
  \bibfield  {author} {\bibinfo {author} {\bibfnamefont {V.}~\bibnamefont
  {Folomeev}},\ }\href {\doibase 10.1103/PhysRevD.97.124009} {\bibfield
  {journal} {\bibinfo  {journal} {Phys. Rev. D}\ }\textbf {\bibinfo {volume}
  {97}},\ \bibinfo {pages} {124009} (\bibinfo {year} {2018})}\BibitemShut
  {NoStop}%
\bibitem [{\citenamefont {Olmo}\ \emph {et~al.}(2020)\citenamefont {Olmo},
  \citenamefont {Rubiera-Garcia},\ and\ \citenamefont {Wojnar}}]{Olmo2020}%
  \BibitemOpen
  \bibfield  {author} {\bibinfo {author} {\bibfnamefont {G.~J.}\ \bibnamefont
  {Olmo}}, \bibinfo {author} {\bibfnamefont {D.}~\bibnamefont
  {Rubiera-Garcia}}, \ and\ \bibinfo {author} {\bibfnamefont {A.}~\bibnamefont
  {Wojnar}},\ }\href {\doibase https://doi.org/10.1016/j.physrep.2020.07.001}
  {\bibfield  {journal} {\bibinfo  {journal} {Physics Reports}\ }\textbf
  {\bibinfo {volume} {876}},\ \bibinfo {pages} {1} (\bibinfo {year}
  {2020})}\BibitemShut {NoStop}%
\bibitem [{\citenamefont {Astashenok}\ \emph {et~al.}(2020)\citenamefont
  {Astashenok} \emph {et~al.}}]{Astashenok2020}%
  \BibitemOpen
  \bibfield  {author} {\bibinfo {author} {\bibfnamefont {A.}~\bibnamefont
  {Astashenok}} \emph {et~al.},\ }\href {\doibase
  https://doi.org/10.1016/j.physletb.2020.135910} {\bibfield  {journal}
  {\bibinfo  {journal} {Physics Letters B}\ }\textbf {\bibinfo {volume}
  {811}},\ \bibinfo {pages} {135910} (\bibinfo {year} {2020})}\BibitemShut
  {NoStop}%
\bibitem [{\citenamefont {Astashenok}\ \emph {et~al.}(2021)\citenamefont
  {Astashenok} \emph {et~al.}}]{Astashenok2021}%
  \BibitemOpen
  \bibfield  {author} {\bibinfo {author} {\bibfnamefont {A.}~\bibnamefont
  {Astashenok}} \emph {et~al.},\ }\href {\doibase
  https://doi.org/10.1016/j.physletb.2021.136222} {\bibfield  {journal}
  {\bibinfo  {journal} {Physics Letters B}\ }\textbf {\bibinfo {volume}
  {816}},\ \bibinfo {pages} {136222} (\bibinfo {year} {2021})}\BibitemShut
  {NoStop}%
\bibitem [{\citenamefont {Numajiri}\ \emph {et~al.}(2022)\citenamefont
  {Numajiri}, \citenamefont {Katsuragawa},\ and\ \citenamefont
  {Nojiri}}]{Numajiri2022}%
  \BibitemOpen
  \bibfield  {author} {\bibinfo {author} {\bibfnamefont {K.}~\bibnamefont
  {Numajiri}}, \bibinfo {author} {\bibfnamefont {T.}~\bibnamefont
  {Katsuragawa}}, \ and\ \bibinfo {author} {\bibfnamefont {S.}~\bibnamefont
  {Nojiri}},\ }\href {\doibase https://doi.org/10.1016/j.physletb.2022.136929}
  {\bibfield  {journal} {\bibinfo  {journal} {Physics Letters B}\ }\textbf
  {\bibinfo {volume} {826}},\ \bibinfo {pages} {136929} (\bibinfo {year}
  {2022})}\BibitemShut {NoStop}%
\bibitem [{\citenamefont {Nobleson}\ \emph {et~al.}(2022)\citenamefont
  {Nobleson}, \citenamefont {Ali},\ and\ \citenamefont {Banik}}]{Nobleson2022}%
  \BibitemOpen
  \bibfield  {author} {\bibinfo {author} {\bibfnamefont {K.}~\bibnamefont
  {Nobleson}}, \bibinfo {author} {\bibfnamefont {A.}~\bibnamefont {Ali}}, \
  and\ \bibinfo {author} {\bibfnamefont {S.}~\bibnamefont {Banik}},\ }\href
  {\doibase https://doi.org/10.1140/epjc/s10052-021-09969-x} {\bibfield
  {journal} {\bibinfo  {journal} {Eur. Phys. J. C}\ }\textbf {\bibinfo {volume}
  {82}},\ \bibinfo {pages} {32} (\bibinfo {year} {2022})}\BibitemShut {NoStop}%
\bibitem [{\citenamefont {Pretel}\ and\ \citenamefont
  {Duarte}(2022)}]{Pretel2022}%
  \BibitemOpen
  \bibfield  {author} {\bibinfo {author} {\bibfnamefont {J.~M.~Z.}\
  \bibnamefont {Pretel}}\ and\ \bibinfo {author} {\bibfnamefont {S.~B.}\
  \bibnamefont {Duarte}},\ }\href {\doibase 10.1088/1361-6382/ac7a88}
  {\bibfield  {journal} {\bibinfo  {journal} {Class. Quantum Grav.}\ }\textbf
  {\bibinfo {volume} {39}},\ \bibinfo {pages} {155003} (\bibinfo {year}
  {2022})}\BibitemShut {NoStop}%
\bibitem [{\citenamefont {Pretel}\ \emph {et~al.}(2022)\citenamefont {Pretel}
  \emph {et~al.}}]{Pretel2022JCAP}%
  \BibitemOpen
  \bibfield  {author} {\bibinfo {author} {\bibfnamefont {J.~M.~Z.}\
  \bibnamefont {Pretel}} \emph {et~al.},\ }\href {\doibase
  10.1088/1475-7516/2022/09/058} {\bibfield  {journal} {\bibinfo  {journal}
  {JCAP}\ }\textbf {\bibinfo {volume} {2022}},\ \bibinfo {pages} {058}
  (\bibinfo {year} {2022})}\BibitemShut {NoStop}%
\bibitem [{\citenamefont {Frieman}\ \emph {et~al.}(2008)\citenamefont
  {Frieman}, \citenamefont {Turner},\ and\ \citenamefont
  {Huterer}}]{Frieman2008}%
  \BibitemOpen
  \bibfield  {author} {\bibinfo {author} {\bibfnamefont {J.~A.}\ \bibnamefont
  {Frieman}}, \bibinfo {author} {\bibfnamefont {M.~S.}\ \bibnamefont {Turner}},
  \ and\ \bibinfo {author} {\bibfnamefont {D.}~\bibnamefont {Huterer}},\ }\href
  {\doibase 10.1146/annurev.astro.46.060407.145243} {\bibfield  {journal}
  {\bibinfo  {journal} {Annu. Rev. Astron. Astrophys.}\ }\textbf {\bibinfo
  {volume} {46}},\ \bibinfo {pages} {385} (\bibinfo {year} {2008})}\BibitemShut
  {NoStop}%
\bibitem [{\citenamefont {Yazadjiev}(2011)}]{Yazadjiev2011}%
  \BibitemOpen
  \bibfield  {author} {\bibinfo {author} {\bibfnamefont {S.~S.}\ \bibnamefont
  {Yazadjiev}},\ }\href {\doibase 10.1103/PhysRevD.83.127501} {\bibfield
  {journal} {\bibinfo  {journal} {Phys. Rev. D}\ }\textbf {\bibinfo {volume}
  {83}},\ \bibinfo {pages} {127501} (\bibinfo {year} {2011})}\BibitemShut
  {NoStop}%
\bibitem [{\citenamefont {Sakti}\ and\ \citenamefont
  {Sulaksono}(2021)}]{Sakti2021}%
  \BibitemOpen
  \bibfield  {author} {\bibinfo {author} {\bibfnamefont {M.~F. A.~R.}\
  \bibnamefont {Sakti}}\ and\ \bibinfo {author} {\bibfnamefont
  {A.}~\bibnamefont {Sulaksono}},\ }\href {\doibase
  10.1103/PhysRevD.103.084042} {\bibfield  {journal} {\bibinfo  {journal}
  {Phys. Rev. D}\ }\textbf {\bibinfo {volume} {103}},\ \bibinfo {pages}
  {084042} (\bibinfo {year} {2021})}\BibitemShut {NoStop}%
\bibitem [{\citenamefont {Smerechynskyi}\ \emph {et~al.}(2021)\citenamefont
  {Smerechynskyi}, \citenamefont {Tsizh},\ and\ \citenamefont
  {Novosyadlyj}}]{Smerechynskyi2021}%
  \BibitemOpen
  \bibfield  {author} {\bibinfo {author} {\bibfnamefont {S.}~\bibnamefont
  {Smerechynskyi}}, \bibinfo {author} {\bibfnamefont {M.}~\bibnamefont
  {Tsizh}}, \ and\ \bibinfo {author} {\bibfnamefont {B.}~\bibnamefont
  {Novosyadlyj}},\ }\href {\doibase 10.1088/1475-7516/2021/02/045} {\bibfield
  {journal} {\bibinfo  {journal} {JCAP}\ }\textbf {\bibinfo {volume} {2021}},\
  \bibinfo {pages} {045} (\bibinfo {year} {2021})}\BibitemShut {NoStop}%
\bibitem [{\citenamefont {Panotopoulos}\ \emph {et~al.}(2021)\citenamefont
  {Panotopoulos}, \citenamefont {Ángel Rincón},\ and\ \citenamefont
  {Lopes}}]{Panotopoulos2021}%
  \BibitemOpen
  \bibfield  {author} {\bibinfo {author} {\bibfnamefont {G.}~\bibnamefont
  {Panotopoulos}}, \bibinfo {author} {\bibnamefont {Ángel Rincón}}, \ and\
  \bibinfo {author} {\bibfnamefont {I.}~\bibnamefont {Lopes}},\ }\href
  {\doibase https://doi.org/10.1016/j.dark.2021.100885} {\bibfield  {journal}
  {\bibinfo  {journal} {Physics of the Dark Universe}\ }\textbf {\bibinfo
  {volume} {34}},\ \bibinfo {pages} {100885} (\bibinfo {year}
  {2021})}\BibitemShut {NoStop}%
\bibitem [{\citenamefont {Bhar}(2021)}]{Bhar2021}%
  \BibitemOpen
  \bibfield  {author} {\bibinfo {author} {\bibfnamefont {P.}~\bibnamefont
  {Bhar}},\ }\href {\doibase https://doi.org/10.1016/j.dark.2021.100879}
  {\bibfield  {journal} {\bibinfo  {journal} {Physics of the Dark Universe}\
  }\textbf {\bibinfo {volume} {34}},\ \bibinfo {pages} {100879} (\bibinfo
  {year} {2021})}\BibitemShut {NoStop}%
\bibitem [{\citenamefont {Chan}\ \emph {et~al.}(2009)\citenamefont {Chan},
  \citenamefont {da~Silva},\ and\ \citenamefont {da~Rocha}}]{Chan2009}%
  \BibitemOpen
  \bibfield  {author} {\bibinfo {author} {\bibfnamefont {R.}~\bibnamefont
  {Chan}}, \bibinfo {author} {\bibfnamefont {M.~F.~A.}\ \bibnamefont
  {da~Silva}}, \ and\ \bibinfo {author} {\bibfnamefont {J.~F.~V.}\ \bibnamefont
  {da~Rocha}},\ }\href {\doibase https://doi.org/10.1007/s10714-008-0755-9}
  {\bibfield  {journal} {\bibinfo  {journal} {Gen. Relativ. Gravit.}\ }\textbf
  {\bibinfo {volume} {41}},\ \bibinfo {pages} {1835} (\bibinfo {year}
  {2009})}\BibitemShut {NoStop}%
\bibitem [{\citenamefont {Rahaman}\ \emph {et~al.}(2010)\citenamefont
  {Rahaman}, \citenamefont {Ray}, \citenamefont {Jafry},\ and\ \citenamefont
  {Chakraborty}}]{Farook2010}%
  \BibitemOpen
  \bibfield  {author} {\bibinfo {author} {\bibfnamefont {F.}~\bibnamefont
  {Rahaman}}, \bibinfo {author} {\bibfnamefont {S.}~\bibnamefont {Ray}},
  \bibinfo {author} {\bibfnamefont {A.~K.}\ \bibnamefont {Jafry}}, \ and\
  \bibinfo {author} {\bibfnamefont {K.}~\bibnamefont {Chakraborty}},\ }\href
  {\doibase 10.1103/PhysRevD.82.104055} {\bibfield  {journal} {\bibinfo
  {journal} {Phys. Rev. D}\ }\textbf {\bibinfo {volume} {82}},\ \bibinfo
  {pages} {104055} (\bibinfo {year} {2010})}\BibitemShut {NoStop}%
\bibitem [{\citenamefont {Ghezzi}(2011)}]{Ghezzi2011}%
  \BibitemOpen
  \bibfield  {author} {\bibinfo {author} {\bibfnamefont {C.~R.}\ \bibnamefont
  {Ghezzi}},\ }\href {\doibase https://doi.org/10.1007/s10509-011-0663-4}
  {\bibfield  {journal} {\bibinfo  {journal} {Astrophys. Space Sci.}\ }\textbf
  {\bibinfo {volume} {333}},\ \bibinfo {pages} {437} (\bibinfo {year}
  {2011})}\BibitemShut {NoStop}%
\bibitem [{\citenamefont {Bhar}\ \emph {et~al.}(2018)\citenamefont {Bhar},
  \citenamefont {Govender},\ and\ \citenamefont {Sharma}}]{Bhar2018}%
  \BibitemOpen
  \bibfield  {author} {\bibinfo {author} {\bibfnamefont {P.}~\bibnamefont
  {Bhar}}, \bibinfo {author} {\bibfnamefont {M.}~\bibnamefont {Govender}}, \
  and\ \bibinfo {author} {\bibfnamefont {R.}~\bibnamefont {Sharma}},\ }\href
  {\doibase https://doi.org/10.1007/s12043-017-1500-2} {\bibfield  {journal}
  {\bibinfo  {journal} {Pramana}\ }\textbf {\bibinfo {volume} {90}},\ \bibinfo
  {pages} {5} (\bibinfo {year} {2018})}\BibitemShut {NoStop}%
\bibitem [{\citenamefont {Tello-Ortiz}\ \emph {et~al.}(2020)\citenamefont
  {Tello-Ortiz} \emph {et~al.}}]{Tello2020}%
  \BibitemOpen
  \bibfield  {author} {\bibinfo {author} {\bibfnamefont {F.}~\bibnamefont
  {Tello-Ortiz}} \emph {et~al.},\ }\href {\doibase
  https://doi.org/10.1140/epjc/s10052-020-7956-0} {\bibfield  {journal}
  {\bibinfo  {journal} {Eur. Phys. J. C}\ }\textbf {\bibinfo {volume} {80}},\
  \bibinfo {pages} {371} (\bibinfo {year} {2020})}\BibitemShut {NoStop}%
\bibitem [{\citenamefont {Estevez-Delgado}\ \emph {et~al.}(2021)\citenamefont
  {Estevez-Delgado} \emph {et~al.}}]{Estevez2021}%
  \BibitemOpen
  \bibfield  {author} {\bibinfo {author} {\bibfnamefont {J.}~\bibnamefont
  {Estevez-Delgado}} \emph {et~al.},\ }\href {\doibase
  https://doi.org/10.1142/S0217732321502138} {\bibfield  {journal} {\bibinfo
  {journal} {Mod. Phys. Lett. A}\ }\textbf {\bibinfo {volume} {36}},\ \bibinfo
  {pages} {2150213} (\bibinfo {year} {2021})}\BibitemShut {NoStop}%
\bibitem [{\citenamefont {Veneroni}\ \emph {et~al.}(2021)\citenamefont
  {Veneroni}, \citenamefont {Braz},\ and\ \citenamefont
  {da~Silva}}]{Veneroni2021}%
  \BibitemOpen
  \bibfield  {author} {\bibinfo {author} {\bibfnamefont {L.~S.~M.}\
  \bibnamefont {Veneroni}}, \bibinfo {author} {\bibfnamefont {A.}~\bibnamefont
  {Braz}}, \ and\ \bibinfo {author} {\bibfnamefont {M.~F.~A.}\ \bibnamefont
  {da~Silva}},\ }\href {\doibase https://doi.org/10.1142/S0218271821500395}
  {\bibfield  {journal} {\bibinfo  {journal} {Int. J. Mod. Phys. D}\ }\textbf
  {\bibinfo {volume} {30}},\ \bibinfo {pages} {2150039} (\bibinfo {year}
  {2021})}\BibitemShut {NoStop}%
\bibitem [{\citenamefont {Grammenos}\ \emph {et~al.}(2021)\citenamefont
  {Grammenos} \emph {et~al.}}]{Grammenos2021}%
  \BibitemOpen
  \bibfield  {author} {\bibinfo {author} {\bibfnamefont {T.}~\bibnamefont
  {Grammenos}} \emph {et~al.},\ }\href {\doibase
  https://doi.org/10.1155/2021/6966689} {\bibfield  {journal} {\bibinfo
  {journal} {Advances in High Energy Physics}\ }\textbf {\bibinfo {volume}
  {2021}},\ \bibinfo {pages} {6966689} (\bibinfo {year} {2021})}\BibitemShut
  {NoStop}%
\bibitem [{\citenamefont {Haghani}\ and\ \citenamefont
  {Harko}(2022)}]{Haghani2022}%
  \BibitemOpen
  \bibfield  {author} {\bibinfo {author} {\bibfnamefont {Z.}~\bibnamefont
  {Haghani}}\ and\ \bibinfo {author} {\bibfnamefont {T.}~\bibnamefont
  {Harko}},\ }\href {\doibase 10.1103/PhysRevD.105.064059} {\bibfield
  {journal} {\bibinfo  {journal} {Phys. Rev. D}\ }\textbf {\bibinfo {volume}
  {105}},\ \bibinfo {pages} {064059} (\bibinfo {year} {2022})}\BibitemShut
  {NoStop}%
\bibitem [{\citenamefont {Bowers}\ and\ \citenamefont
  {Liang}(1974)}]{BowersLiang1974}%
  \BibitemOpen
  \bibfield  {author} {\bibinfo {author} {\bibfnamefont {R.~L.}\ \bibnamefont
  {Bowers}}\ and\ \bibinfo {author} {\bibfnamefont {E.~P.~T.}\ \bibnamefont
  {Liang}},\ }\href@noop {} {\bibfield  {journal} {\bibinfo  {journal}
  {Astrophys. J.}\ }\textbf {\bibinfo {volume} {188}},\ \bibinfo {pages} {657}
  (\bibinfo {year} {1974})}\BibitemShut {NoStop}%
\bibitem [{\citenamefont {Cosenza}\ \emph {et~al.}(1981)\citenamefont
  {Cosenza}, \citenamefont {Herrera}, \citenamefont {Esculpi},\ and\
  \citenamefont {Witten}}]{CHEW1981}%
  \BibitemOpen
  \bibfield  {author} {\bibinfo {author} {\bibfnamefont {M.}~\bibnamefont
  {Cosenza}}, \bibinfo {author} {\bibfnamefont {L.}~\bibnamefont {Herrera}},
  \bibinfo {author} {\bibfnamefont {M.}~\bibnamefont {Esculpi}}, \ and\
  \bibinfo {author} {\bibfnamefont {L.}~\bibnamefont {Witten}},\ }\href
  {\doibase https://doi.org/10.1063/1.524742} {\bibfield  {journal} {\bibinfo
  {journal} {J. Math. Phys.}\ }\textbf {\bibinfo {volume} {22}},\ \bibinfo
  {pages} {118} (\bibinfo {year} {1981})}\BibitemShut {NoStop}%
\bibitem [{\citenamefont {Horvat}\ \emph {et~al.}(2010)\citenamefont {Horvat},
  \citenamefont {Iliji{\'{c}}},\ and\ \citenamefont
  {Marunovi{\'{c}}}}]{Horvat2011}%
  \BibitemOpen
  \bibfield  {author} {\bibinfo {author} {\bibfnamefont {D.}~\bibnamefont
  {Horvat}}, \bibinfo {author} {\bibfnamefont {S.}~\bibnamefont
  {Iliji{\'{c}}}}, \ and\ \bibinfo {author} {\bibfnamefont {A.}~\bibnamefont
  {Marunovi{\'{c}}}},\ }\href {\doibase 10.1088/0264-9381/28/2/025009}
  {\bibfield  {journal} {\bibinfo  {journal} {Class. Quantum Grav.}\ }\textbf
  {\bibinfo {volume} {28}},\ \bibinfo {pages} {025009} (\bibinfo {year}
  {2010})}\BibitemShut {NoStop}%
\bibitem [{\citenamefont {Doneva}\ and\ \citenamefont
  {Yazadjiev}(2012)}]{Doneva2012}%
  \BibitemOpen
  \bibfield  {author} {\bibinfo {author} {\bibfnamefont {D.~D.}\ \bibnamefont
  {Doneva}}\ and\ \bibinfo {author} {\bibfnamefont {S.~S.}\ \bibnamefont
  {Yazadjiev}},\ }\href {\doibase 10.1103/PhysRevD.85.124023} {\bibfield
  {journal} {\bibinfo  {journal} {Phys. Rev. D}\ }\textbf {\bibinfo {volume}
  {85}},\ \bibinfo {pages} {124023} (\bibinfo {year} {2012})}\BibitemShut
  {NoStop}%
\bibitem [{\citenamefont {Herrera}\ and\ \citenamefont
  {Barreto}(2013)}]{HerreraBarreto2013}%
  \BibitemOpen
  \bibfield  {author} {\bibinfo {author} {\bibfnamefont {L.}~\bibnamefont
  {Herrera}}\ and\ \bibinfo {author} {\bibfnamefont {W.}~\bibnamefont
  {Barreto}},\ }\href {\doibase 10.1103/PhysRevD.88.084022} {\bibfield
  {journal} {\bibinfo  {journal} {Phys. Rev. D}\ }\textbf {\bibinfo {volume}
  {88}},\ \bibinfo {pages} {084022} (\bibinfo {year} {2013})}\BibitemShut
  {NoStop}%
\bibitem [{\citenamefont {Raposo}\ \emph {et~al.}(2019)\citenamefont {Raposo}
  \emph {et~al.}}]{Raposo2019}%
  \BibitemOpen
  \bibfield  {author} {\bibinfo {author} {\bibfnamefont {G.}~\bibnamefont
  {Raposo}} \emph {et~al.},\ }\href {\doibase 10.1103/PhysRevD.99.104072}
  {\bibfield  {journal} {\bibinfo  {journal} {Phys. Rev. D}\ }\textbf {\bibinfo
  {volume} {99}},\ \bibinfo {pages} {104072} (\bibinfo {year}
  {2019})}\BibitemShut {NoStop}%
\bibitem [{\citenamefont {Pretel}(2020)}]{Pretel2020EPJC}%
  \BibitemOpen
  \bibfield  {author} {\bibinfo {author} {\bibfnamefont {J.~M.~Z.}\
  \bibnamefont {Pretel}},\ }\href {\doibase
  https://doi.org/10.1140/epjc/s10052-020-8301-3} {\bibfield  {journal}
  {\bibinfo  {journal} {Eur. Phys. J. C}\ }\textbf {\bibinfo {volume} {80}},\
  \bibinfo {pages} {726} (\bibinfo {year} {2020})}\BibitemShut {NoStop}%
\bibitem [{\citenamefont {Rizaldy}\ \emph {et~al.}(2019)\citenamefont
  {Rizaldy}, \citenamefont {Alfarasyi}, \citenamefont {Sulaksono},\ and\
  \citenamefont {Sumaryada}}]{Rizaldy2019}%
  \BibitemOpen
  \bibfield  {author} {\bibinfo {author} {\bibfnamefont {R.}~\bibnamefont
  {Rizaldy}}, \bibinfo {author} {\bibfnamefont {A.~R.}\ \bibnamefont
  {Alfarasyi}}, \bibinfo {author} {\bibfnamefont {A.}~\bibnamefont
  {Sulaksono}}, \ and\ \bibinfo {author} {\bibfnamefont {T.}~\bibnamefont
  {Sumaryada}},\ }\href {\doibase 10.1103/PhysRevC.100.055804} {\bibfield
  {journal} {\bibinfo  {journal} {Phys. Rev. C}\ }\textbf {\bibinfo {volume}
  {100}},\ \bibinfo {pages} {055804} (\bibinfo {year} {2019})}\BibitemShut
  {NoStop}%
\bibitem [{\citenamefont {Becerra-Vergara}\ \emph {et~al.}(2019)\citenamefont
  {Becerra-Vergara}, \citenamefont {Mojica}, \citenamefont {Lora-Clavijo},\
  and\ \citenamefont {Cruz-Osorio}}]{Becerra2019}%
  \BibitemOpen
  \bibfield  {author} {\bibinfo {author} {\bibfnamefont {E.~A.}\ \bibnamefont
  {Becerra-Vergara}}, \bibinfo {author} {\bibfnamefont {S.}~\bibnamefont
  {Mojica}}, \bibinfo {author} {\bibfnamefont {F.~D.}\ \bibnamefont
  {Lora-Clavijo}}, \ and\ \bibinfo {author} {\bibfnamefont {A.}~\bibnamefont
  {Cruz-Osorio}},\ }\href {\doibase 10.1103/PhysRevD.100.103006} {\bibfield
  {journal} {\bibinfo  {journal} {Phys. Rev. D}\ }\textbf {\bibinfo {volume}
  {100}},\ \bibinfo {pages} {103006} (\bibinfo {year} {2019})}\BibitemShut
  {NoStop}%
\bibitem [{\citenamefont {Danarianto}\ and\ \citenamefont
  {Sulaksono}(2019)}]{Danarianto2019}%
  \BibitemOpen
  \bibfield  {author} {\bibinfo {author} {\bibfnamefont {M.~D.}\ \bibnamefont
  {Danarianto}}\ and\ \bibinfo {author} {\bibfnamefont {A.}~\bibnamefont
  {Sulaksono}},\ }\href {\doibase 10.1103/PhysRevD.100.064042} {\bibfield
  {journal} {\bibinfo  {journal} {Phys. Rev. D}\ }\textbf {\bibinfo {volume}
  {100}},\ \bibinfo {pages} {064042} (\bibinfo {year} {2019})}\BibitemShut
  {NoStop}%
\bibitem [{\citenamefont {{Hartle}}(1967)}]{Hartle1967}%
  \BibitemOpen
  \bibfield  {author} {\bibinfo {author} {\bibfnamefont {J.~B.}\ \bibnamefont
  {{Hartle}}},\ }\href {\doibase 10.1086/149400} {\bibfield  {journal}
  {\bibinfo  {journal} {\apj}\ }\textbf {\bibinfo {volume} {150}},\ \bibinfo
  {pages} {1005} (\bibinfo {year} {1967})}\BibitemShut {NoStop}%
\bibitem [{\citenamefont {Hartle}(1973)}]{Hartle1973}%
  \BibitemOpen
  \bibfield  {author} {\bibinfo {author} {\bibfnamefont {J.~B.}\ \bibnamefont
  {Hartle}},\ }\href {\doibase https://doi.org/10.1007/BF02637163} {\bibfield
  {journal} {\bibinfo  {journal} {Astrophys. Space Sci.}\ }\textbf {\bibinfo
  {volume} {24}},\ \bibinfo {pages} {385} (\bibinfo {year} {1973})}\BibitemShut
  {NoStop}%
\bibitem [{\citenamefont {Glendenning}(2000)}]{Glendenning}%
  \BibitemOpen
  \bibfield  {author} {\bibinfo {author} {\bibfnamefont {N.~K.}\ \bibnamefont
  {Glendenning}},\ }\href@noop {} {\emph {\bibinfo {title} {Compact Stars:
  Nuclear Physics, Particle Physics, and General Relativity}}},\ \bibinfo
  {edition} {2nd}\ ed.\ (\bibinfo  {publisher} {Astron. Astrophys. Library,
  Springer},\ \bibinfo {address} {New York},\ \bibinfo {year}
  {2000})\BibitemShut {NoStop}%
\bibitem [{\citenamefont {Most}\ \emph {et~al.}(2018)\citenamefont {Most},
  \citenamefont {Weih}, \citenamefont {Rezzolla},\ and\ \citenamefont
  {Schaffner-Bielich}}]{Most2018}%
  \BibitemOpen
  \bibfield  {author} {\bibinfo {author} {\bibfnamefont {E.~R.}\ \bibnamefont
  {Most}}, \bibinfo {author} {\bibfnamefont {L.~R.}\ \bibnamefont {Weih}},
  \bibinfo {author} {\bibfnamefont {L.}~\bibnamefont {Rezzolla}}, \ and\
  \bibinfo {author} {\bibfnamefont {J.}~\bibnamefont {Schaffner-Bielich}},\
  }\href {\doibase 10.1103/PhysRevLett.120.261103} {\bibfield  {journal}
  {\bibinfo  {journal} {Phys. Rev. Lett.}\ }\textbf {\bibinfo {volume} {120}},\
  \bibinfo {pages} {261103} (\bibinfo {year} {2018})}\BibitemShut {NoStop}%
\bibitem [{\citenamefont {Chatziioannou}(2020)}]{Chatziioannou2020}%
  \BibitemOpen
  \bibfield  {author} {\bibinfo {author} {\bibfnamefont {K.}~\bibnamefont
  {Chatziioannou}},\ }\href {\doibase
  https://doi.org/10.1007/s10714-020-02754-3} {\bibfield  {journal} {\bibinfo
  {journal} {Gen. Relativ. Gravit.}\ }\textbf {\bibinfo {volume} {52}},\
  \bibinfo {pages} {109} (\bibinfo {year} {2020})}\BibitemShut {NoStop}%
\bibitem [{\citenamefont {Hinderer}(2008)}]{Hinderer2008}%
  \BibitemOpen
  \bibfield  {author} {\bibinfo {author} {\bibfnamefont {T.}~\bibnamefont
  {Hinderer}},\ }\href {https://doi.org/10.1086/533487} {\bibfield  {journal}
  {\bibinfo  {journal} {Astrophys. J.}\ }\textbf {\bibinfo {volume} {677}},\
  \bibinfo {pages} {1216} (\bibinfo {year} {2008})}\BibitemShut {NoStop}%
\bibitem [{\citenamefont {Damour}\ and\ \citenamefont
  {Nagar}(2009)}]{Damour2009}%
  \BibitemOpen
  \bibfield  {author} {\bibinfo {author} {\bibfnamefont {T.}~\bibnamefont
  {Damour}}\ and\ \bibinfo {author} {\bibfnamefont {A.}~\bibnamefont {Nagar}},\
  }\href {\doibase 10.1103/PhysRevD.80.084035} {\bibfield  {journal} {\bibinfo
  {journal} {Phys. Rev. D}\ }\textbf {\bibinfo {volume} {80}},\ \bibinfo
  {pages} {084035} (\bibinfo {year} {2009})}\BibitemShut {NoStop}%
\bibitem [{\citenamefont {Binnington}\ and\ \citenamefont
  {Poisson}(2009)}]{Binnington2009}%
  \BibitemOpen
  \bibfield  {author} {\bibinfo {author} {\bibfnamefont {T.}~\bibnamefont
  {Binnington}}\ and\ \bibinfo {author} {\bibfnamefont {E.}~\bibnamefont
  {Poisson}},\ }\href {\doibase 10.1103/PhysRevD.80.084018} {\bibfield
  {journal} {\bibinfo  {journal} {Phys. Rev. D}\ }\textbf {\bibinfo {volume}
  {80}},\ \bibinfo {pages} {084018} (\bibinfo {year} {2009})}\BibitemShut
  {NoStop}%
\bibitem [{\citenamefont {Postnikov}\ \emph {et~al.}(2010)\citenamefont
  {Postnikov}, \citenamefont {Prakash},\ and\ \citenamefont
  {Lattimer}}]{Postnikov2021}%
  \BibitemOpen
  \bibfield  {author} {\bibinfo {author} {\bibfnamefont {S.}~\bibnamefont
  {Postnikov}}, \bibinfo {author} {\bibfnamefont {M.}~\bibnamefont {Prakash}},
  \ and\ \bibinfo {author} {\bibfnamefont {J.~M.}\ \bibnamefont {Lattimer}},\
  }\href {\doibase 10.1103/PhysRevD.82.024016} {\bibfield  {journal} {\bibinfo
  {journal} {Phys. Rev. D}\ }\textbf {\bibinfo {volume} {82}},\ \bibinfo
  {pages} {024016} (\bibinfo {year} {2010})}\BibitemShut {NoStop}%
\bibitem [{\citenamefont {Chaves}\ and\ \citenamefont
  {Hinderer}(2019)}]{Chaves2019}%
  \BibitemOpen
  \bibfield  {author} {\bibinfo {author} {\bibfnamefont {A.~G.}\ \bibnamefont
  {Chaves}}\ and\ \bibinfo {author} {\bibfnamefont {T.}~\bibnamefont
  {Hinderer}},\ }\href {\doibase 10.1088/1361-6471/ab45be} {\bibfield
  {journal} {\bibinfo  {journal} {J. Phys. G: Nucl. Part. Phys.}\ }\textbf
  {\bibinfo {volume} {46}},\ \bibinfo {pages} {123002} (\bibinfo {year}
  {2019})}\BibitemShut {NoStop}%
\bibitem [{\citenamefont {Dietrich}\ \emph {et~al.}(2021)\citenamefont
  {Dietrich}, \citenamefont {Hinderer},\ and\ \citenamefont
  {Samajdar}}]{Dietrich2021}%
  \BibitemOpen
  \bibfield  {author} {\bibinfo {author} {\bibfnamefont {T.}~\bibnamefont
  {Dietrich}}, \bibinfo {author} {\bibfnamefont {T.}~\bibnamefont {Hinderer}},
  \ and\ \bibinfo {author} {\bibfnamefont {A.}~\bibnamefont {Samajdar}},\
  }\href {\doibase https://doi.org/10.1007/s10714-020-02751-6} {\bibfield
  {journal} {\bibinfo  {journal} {Gen. Relativ. Gravit.}\ }\textbf {\bibinfo
  {volume} {53}},\ \bibinfo {pages} {27} (\bibinfo {year} {2021})}\BibitemShut
  {NoStop}%
\bibitem [{\citenamefont {Kumari}\ and\ \citenamefont
  {Kumar}(2021)}]{Kumari2021}%
  \BibitemOpen
  \bibfield  {author} {\bibinfo {author} {\bibfnamefont {M.}~\bibnamefont
  {Kumari}}\ and\ \bibinfo {author} {\bibfnamefont {A.}~\bibnamefont {Kumar}},\
  }\href {\doibase https://doi.org/10.1140/epjc/s10052-021-09576-w} {\bibfield
  {journal} {\bibinfo  {journal} {Eur. Phys. J. C}\ }\textbf {\bibinfo {volume}
  {81}},\ \bibinfo {pages} {791} (\bibinfo {year} {2021})}\BibitemShut
  {NoStop}%
\bibitem [{\citenamefont {Regge}\ and\ \citenamefont
  {Wheeler}(1957)}]{Regge1957}%
  \BibitemOpen
  \bibfield  {author} {\bibinfo {author} {\bibfnamefont {T.}~\bibnamefont
  {Regge}}\ and\ \bibinfo {author} {\bibfnamefont {J.~A.}\ \bibnamefont
  {Wheeler}},\ }\href {\doibase 10.1103/PhysRev.108.1063} {\bibfield  {journal}
  {\bibinfo  {journal} {Phys. Rev.}\ }\textbf {\bibinfo {volume} {108}},\
  \bibinfo {pages} {1063} (\bibinfo {year} {1957})}\BibitemShut {NoStop}%
\bibitem [{\citenamefont {Biswas}\ and\ \citenamefont
  {Bose}(2019)}]{Biswas2019}%
  \BibitemOpen
  \bibfield  {author} {\bibinfo {author} {\bibfnamefont {B.}~\bibnamefont
  {Biswas}}\ and\ \bibinfo {author} {\bibfnamefont {S.}~\bibnamefont {Bose}},\
  }\href {\doibase 10.1103/PhysRevD.99.104002} {\bibfield  {journal} {\bibinfo
  {journal} {Phys. Rev. D}\ }\textbf {\bibinfo {volume} {99}},\ \bibinfo
  {pages} {104002} (\bibinfo {year} {2019})}\BibitemShut {NoStop}%
\bibitem [{\citenamefont {Cunha}\ \emph {et~al.}(2004)\citenamefont {Cunha},
  \citenamefont {Alcaniz},\ and\ \citenamefont {Lima}}]{Cunha2004}%
  \BibitemOpen
  \bibfield  {author} {\bibinfo {author} {\bibfnamefont {J.~V.}\ \bibnamefont
  {Cunha}}, \bibinfo {author} {\bibfnamefont {J.~S.}\ \bibnamefont {Alcaniz}},
  \ and\ \bibinfo {author} {\bibfnamefont {J.~A.~S.}\ \bibnamefont {Lima}},\
  }\href {\doibase 10.1103/PhysRevD.69.083501} {\bibfield  {journal} {\bibinfo
  {journal} {Phys. Rev. D}\ }\textbf {\bibinfo {volume} {69}},\ \bibinfo
  {pages} {083501} (\bibinfo {year} {2004})}\BibitemShut {NoStop}%
\bibitem [{\citenamefont {Gorini}\ \emph {et~al.}(2008)\citenamefont {Gorini}
  \emph {et~al.}}]{Gorini2008}%
  \BibitemOpen
  \bibfield  {author} {\bibinfo {author} {\bibfnamefont {V.}~\bibnamefont
  {Gorini}} \emph {et~al.},\ }\href {\doibase 10.1088/1475-7516/2008/02/016}
  {\bibfield  {journal} {\bibinfo  {journal} {JCAP}\ }\textbf {\bibinfo
  {volume} {2008}},\ \bibinfo {pages} {016} (\bibinfo {year}
  {2008})}\BibitemShut {NoStop}%
\bibitem [{\citenamefont {Piattella}(2010)}]{Piattella2010}%
  \BibitemOpen
  \bibfield  {author} {\bibinfo {author} {\bibfnamefont {O.~F.}\ \bibnamefont
  {Piattella}},\ }\href {\doibase 10.1088/1475-7516/2010/03/012} {\bibfield
  {journal} {\bibinfo  {journal} {JCAP}\ }\textbf {\bibinfo {volume} {2010}},\
  \bibinfo {pages} {012} (\bibinfo {year} {2010})}\BibitemShut {NoStop}%
\bibitem [{\citenamefont {Salahedin}\ \emph {et~al.}(2022)\citenamefont
  {Salahedin} \emph {et~al.}}]{Salahedin2022}%
  \BibitemOpen
  \bibfield  {author} {\bibinfo {author} {\bibfnamefont {S.~F.}\ \bibnamefont
  {Salahedin}} \emph {et~al.},\ }\href {\doibase
  https://doi.org/10.1007/s12036-022-09797-9} {\bibfield  {journal} {\bibinfo
  {journal} {J. Astrophys. Astron.}\ }\textbf {\bibinfo {volume} {43}},\
  \bibinfo {pages} {14} (\bibinfo {year} {2022})}\BibitemShut {NoStop}%
\bibitem [{\citenamefont {von Marttens}\ \emph {et~al.}(2022)\citenamefont {von
  Marttens}, \citenamefont {Barbosa},\ and\ \citenamefont
  {Alcaniz}}]{Marttens2022}%
  \BibitemOpen
  \bibfield  {author} {\bibinfo {author} {\bibfnamefont {R.}~\bibnamefont {von
  Marttens}}, \bibinfo {author} {\bibfnamefont {D.}~\bibnamefont {Barbosa}}, \
  and\ \bibinfo {author} {\bibfnamefont {J.}~\bibnamefont {Alcaniz}},\ }\href
  {https://arxiv.org/abs/2208.06302} {\bibfield  {journal} {\bibinfo  {journal}
  {arXiv:2208.06302 [astro-ph.CO]}\ } (\bibinfo {year} {2022})}\BibitemShut
  {NoStop}%
\bibitem [{\citenamefont {Silva}\ \emph {et~al.}(2015)\citenamefont {Silva}
  \emph {et~al.}}]{Silva2015}%
  \BibitemOpen
  \bibfield  {author} {\bibinfo {author} {\bibfnamefont {H.~O.}\ \bibnamefont
  {Silva}} \emph {et~al.},\ }\href {\doibase 10.1088/0264-9381/32/14/145008}
  {\bibfield  {journal} {\bibinfo  {journal} {Class. Quantum Grav.}\ }\textbf
  {\bibinfo {volume} {32}},\ \bibinfo {pages} {145008} (\bibinfo {year}
  {2015})}\BibitemShut {NoStop}%
\bibitem [{\citenamefont {Yagi}\ and\ \citenamefont {Yunes}(2015)}]{Yagi2015}%
  \BibitemOpen
  \bibfield  {author} {\bibinfo {author} {\bibfnamefont {K.}~\bibnamefont
  {Yagi}}\ and\ \bibinfo {author} {\bibfnamefont {N.}~\bibnamefont {Yunes}},\
  }\href {\doibase 10.1103/PhysRevD.91.123008} {\bibfield  {journal} {\bibinfo
  {journal} {Phys. Rev. D}\ }\textbf {\bibinfo {volume} {91}},\ \bibinfo
  {pages} {123008} (\bibinfo {year} {2015})}\BibitemShut {NoStop}%
\bibitem [{\citenamefont {Rahmansyah}\ \emph {et~al.}(2020)\citenamefont
  {Rahmansyah} \emph {et~al.}}]{Rahmansyah2020}%
  \BibitemOpen
  \bibfield  {author} {\bibinfo {author} {\bibfnamefont {A.}~\bibnamefont
  {Rahmansyah}} \emph {et~al.},\ }\href {\doibase
  https://doi.org/10.1140/epjc/s10052-020-8361-4} {\bibfield  {journal}
  {\bibinfo  {journal} {Eur. Phys. J. C}\ }\textbf {\bibinfo {volume} {80}},\
  \bibinfo {pages} {769} (\bibinfo {year} {2020})}\BibitemShut {NoStop}%
\bibitem [{\citenamefont {Rahmansyah}\ and\ \citenamefont
  {Sulaksono}(2021)}]{Rahmansyah2021}%
  \BibitemOpen
  \bibfield  {author} {\bibinfo {author} {\bibfnamefont {A.}~\bibnamefont
  {Rahmansyah}}\ and\ \bibinfo {author} {\bibfnamefont {A.}~\bibnamefont
  {Sulaksono}},\ }\href {\doibase 10.1103/PhysRevC.104.065805} {\bibfield
  {journal} {\bibinfo  {journal} {Phys. Rev. C}\ }\textbf {\bibinfo {volume}
  {104}},\ \bibinfo {pages} {065805} (\bibinfo {year} {2021})}\BibitemShut
  {NoStop}%
\bibitem [{\citenamefont {Pretel}(2022)}]{PretelMPLA2022}%
  \BibitemOpen
  \bibfield  {author} {\bibinfo {author} {\bibfnamefont {J.~M.~Z.}\
  \bibnamefont {Pretel}},\ }\href {\doibase
  https://doi.org/10.1142/S0217732322501887} {\bibfield  {journal} {\bibinfo
  {journal} {Mod. Phys. Lett. A}\ }\textbf {\bibinfo {volume} {37}},\ \bibinfo
  {pages} {2250188} (\bibinfo {year} {2022})}\BibitemShut {NoStop}%
\bibitem [{\citenamefont {Kumar}\ and\ \citenamefont
  {Bharti}(2022)}]{Kumar2022}%
  \BibitemOpen
  \bibfield  {author} {\bibinfo {author} {\bibfnamefont {J.}~\bibnamefont
  {Kumar}}\ and\ \bibinfo {author} {\bibfnamefont {P.}~\bibnamefont {Bharti}},\
  }\href {\doibase https://doi.org/10.1016/j.newar.2022.101662} {\bibfield
  {journal} {\bibinfo  {journal} {New Astronomy Reviews}\ }\textbf {\bibinfo
  {volume} {95}},\ \bibinfo {pages} {101662} (\bibinfo {year}
  {2022})}\BibitemShut {NoStop}%
\bibitem [{\citenamefont {Annala}\ \emph {et~al.}(2020)\citenamefont {Annala}
  \emph {et~al.}}]{Annala2020}%
  \BibitemOpen
  \bibfield  {author} {\bibinfo {author} {\bibfnamefont {E.}~\bibnamefont
  {Annala}} \emph {et~al.},\ }\href {\doibase
  https://doi.org/10.1038/s41567-020-0914-9} {\bibfield  {journal} {\bibinfo
  {journal} {Nature Phys.}\ }\textbf {\bibinfo {volume} {16}},\ \bibinfo
  {pages} {907} (\bibinfo {year} {2020})}\BibitemShut {NoStop}%
\bibitem [{\citenamefont {Demorest}\ \emph {et~al.}(2010)\citenamefont
  {Demorest}, \citenamefont {Pennucci}, \citenamefont {Ransom}, \citenamefont
  {Roberts},\ and\ \citenamefont {Hessels}}]{Demorest:2010bx}%
  \BibitemOpen
  \bibfield  {author} {\bibinfo {author} {\bibfnamefont {P.}~\bibnamefont
  {Demorest}}, \bibinfo {author} {\bibfnamefont {T.}~\bibnamefont {Pennucci}},
  \bibinfo {author} {\bibfnamefont {S.}~\bibnamefont {Ransom}}, \bibinfo
  {author} {\bibfnamefont {M.}~\bibnamefont {Roberts}}, \ and\ \bibinfo
  {author} {\bibfnamefont {J.}~\bibnamefont {Hessels}},\ }\href {\doibase
  10.1038/nature09466} {\bibfield  {journal} {\bibinfo  {journal} {Nature}\
  }\textbf {\bibinfo {volume} {467}},\ \bibinfo {pages} {1081} (\bibinfo {year}
  {2010})}\BibitemShut {NoStop}%
\bibitem [{\citenamefont {Antoniadis}\ \emph {et~al.}(2013)\citenamefont
  {Antoniadis} \emph {et~al.}}]{Antoniadis:2013pzd}%
  \BibitemOpen
  \bibfield  {author} {\bibinfo {author} {\bibfnamefont {J.}~\bibnamefont
  {Antoniadis}} \emph {et~al.},\ }\href {\doibase 10.1126/science.1233232}
  {\bibfield  {journal} {\bibinfo  {journal} {Science}\ }\textbf {\bibinfo
  {volume} {340}},\ \bibinfo {pages} {6131} (\bibinfo {year}
  {2013})}\BibitemShut {NoStop}%
\bibitem [{\citenamefont {Cromartie}\ \emph {et~al.}(2019)\citenamefont
  {Cromartie} \emph {et~al.}}]{Cromartie2019}%
  \BibitemOpen
  \bibfield  {author} {\bibinfo {author} {\bibfnamefont {H.~T.}\ \bibnamefont
  {Cromartie}} \emph {et~al.},\ }\href {\doibase
  https://doi.org/10.1038/s41550-019-0880-2} {\bibfield  {journal} {\bibinfo
  {journal} {Nature Astronomy}\ }\textbf {\bibinfo {volume} {4}},\ \bibinfo
  {pages} {72} (\bibinfo {year} {2019})}\BibitemShut {NoStop}%
\bibitem [{\citenamefont {Miller}\ \emph
  {et~al.}(2019{\natexlab{a}})\citenamefont {Miller} \emph
  {et~al.}}]{Miller2019}%
  \BibitemOpen
  \bibfield  {author} {\bibinfo {author} {\bibfnamefont {M.~C.}\ \bibnamefont
  {Miller}} \emph {et~al.},\ }\href {\doibase 10.3847/2041-8213/ab50c5}
  {\bibfield  {journal} {\bibinfo  {journal} {Astrophys. J. Lett.}\ }\textbf
  {\bibinfo {volume} {887}},\ \bibinfo {pages} {L24} (\bibinfo {year}
  {2019}{\natexlab{a}})}\BibitemShut {NoStop}%
\bibitem [{\citenamefont {Riley}\ \emph
  {et~al.}(2019{\natexlab{a}})\citenamefont {Riley} \emph
  {et~al.}}]{Riley2019}%
  \BibitemOpen
  \bibfield  {author} {\bibinfo {author} {\bibfnamefont {T.~E.}\ \bibnamefont
  {Riley}} \emph {et~al.},\ }\href {\doibase 10.3847/2041-8213/ab481c}
  {\bibfield  {journal} {\bibinfo  {journal} {Astrophys. J. Lett.}\ }\textbf
  {\bibinfo {volume} {887}},\ \bibinfo {pages} {L21} (\bibinfo {year}
  {2019}{\natexlab{a}})}\BibitemShut {NoStop}%
\bibitem [{\citenamefont {Raaijmakers}\ \emph {et~al.}(2019)\citenamefont
  {Raaijmakers} \emph {et~al.}}]{Raaijmakers2019}%
  \BibitemOpen
  \bibfield  {author} {\bibinfo {author} {\bibfnamefont {G.}~\bibnamefont
  {Raaijmakers}} \emph {et~al.},\ }\href {\doibase 10.3847/2041-8213/ab451a}
  {\bibfield  {journal} {\bibinfo  {journal} {Astrophys. J. Lett.}\ }\textbf
  {\bibinfo {volume} {887}},\ \bibinfo {pages} {L22} (\bibinfo {year}
  {2019})}\BibitemShut {NoStop}%
\bibitem [{\citenamefont {Abbott}\ \emph {et~al.}(2020)\citenamefont {Abbott}
  \emph {et~al.}}]{Abbott2020}%
  \BibitemOpen
  \bibfield  {author} {\bibinfo {author} {\bibfnamefont {R.}~\bibnamefont
  {Abbott}} \emph {et~al.},\ }\href {\doibase
  https://doi.org/10.3847/2041-8213/ab960f} {\bibfield  {journal} {\bibinfo
  {journal} {Astrophys. J. Lett.}\ }\textbf {\bibinfo {volume} {896}},\
  \bibinfo {pages} {L44} (\bibinfo {year} {2020})}\BibitemShut {NoStop}%
\bibitem [{\citenamefont {Iyer}\ \emph {et~al.}(1985)\citenamefont {Iyer},
  \citenamefont {Vishveshwara},\ and\ \citenamefont {Dhurandhar}}]{Iyer1985}%
  \BibitemOpen
  \bibfield  {author} {\bibinfo {author} {\bibfnamefont {B.~R.}\ \bibnamefont
  {Iyer}}, \bibinfo {author} {\bibfnamefont {C.~V.}\ \bibnamefont
  {Vishveshwara}}, \ and\ \bibinfo {author} {\bibfnamefont {S.~V.}\
  \bibnamefont {Dhurandhar}},\ }\href {\doibase 10.1088/0264-9381/2/2/013}
  {\bibfield  {journal} {\bibinfo  {journal} {Class. Quantum Grav.}\ }\textbf
  {\bibinfo {volume} {2}},\ \bibinfo {pages} {219} (\bibinfo {year}
  {1985})}\BibitemShut {NoStop}%
\bibitem [{\citenamefont {Douchin}\ and\ \citenamefont
  {Haensel}(2001)}]{DouchinHaensel2001}%
  \BibitemOpen
  \bibfield  {author} {\bibinfo {author} {\bibfnamefont {F.}~\bibnamefont
  {Douchin}}\ and\ \bibinfo {author} {\bibfnamefont {P.}~\bibnamefont
  {Haensel}},\ }\href {\doibase 10.1051/0004-6361:20011402} {\bibfield
  {journal} {\bibinfo  {journal} {A\&A}\ }\textbf {\bibinfo {volume} {380}},\
  \bibinfo {pages} {151} (\bibinfo {year} {2001})}\BibitemShut {NoStop}%
\bibitem [{\citenamefont {Linares}\ \emph {et~al.}(2018)\citenamefont
  {Linares}, \citenamefont {Shahbaz},\ and\ \citenamefont
  {Casares}}]{Linares2018}%
  \BibitemOpen
  \bibfield  {author} {\bibinfo {author} {\bibfnamefont {M.}~\bibnamefont
  {Linares}}, \bibinfo {author} {\bibfnamefont {T.}~\bibnamefont {Shahbaz}}, \
  and\ \bibinfo {author} {\bibfnamefont {J.}~\bibnamefont {Casares}},\ }\href
  {\doibase https://doi.org/10.3847/1538-4357/aabde6} {\bibfield  {journal}
  {\bibinfo  {journal} {Astrophys. J.}\ }\textbf {\bibinfo {volume} {859}},\
  \bibinfo {pages} {54} (\bibinfo {year} {2018})}\BibitemShut {NoStop}%
\bibitem [{\citenamefont {Miller}\ \emph
  {et~al.}(2019{\natexlab{b}})\citenamefont {Miller} \emph
  {et~al.}}]{Miller:2019cac}%
  \BibitemOpen
  \bibfield  {author} {\bibinfo {author} {\bibfnamefont {M.~C.}\ \bibnamefont
  {Miller}} \emph {et~al.},\ }\href {\doibase 10.3847/2041-8213/ab50c5}
  {\bibfield  {journal} {\bibinfo  {journal} {Astrophys. J. Lett.}\ }\textbf
  {\bibinfo {volume} {887}},\ \bibinfo {pages} {L24} (\bibinfo {year}
  {2019}{\natexlab{b}})}\BibitemShut {NoStop}%
\bibitem [{\citenamefont {Riley}\ \emph
  {et~al.}(2019{\natexlab{b}})\citenamefont {Riley} \emph
  {et~al.}}]{Riley:2019yda}%
  \BibitemOpen
  \bibfield  {author} {\bibinfo {author} {\bibfnamefont {T.~E.}\ \bibnamefont
  {Riley}} \emph {et~al.},\ }\href {\doibase 10.3847/2041-8213/ab481c}
  {\bibfield  {journal} {\bibinfo  {journal} {Astrophys. J. Lett.}\ }\textbf
  {\bibinfo {volume} {887}},\ \bibinfo {pages} {L21} (\bibinfo {year}
  {2019}{\natexlab{b}})}\BibitemShut {NoStop}%
\bibitem [{\citenamefont {Pretel}\ and\ \citenamefont
  {da Silva}(2020)}]{Pretel2020MNRAS}%
  \BibitemOpen
  \bibfield  {author} {\bibinfo {author} {\bibfnamefont {J.~M.~Z.}\
  \bibnamefont {Pretel}}\ and\ \bibinfo {author} {\bibfnamefont {M.~F.~A.}\
  \bibnamefont {da Silva}},\ }\href {\doibase 10.1093/mnras/staa1493}
  {\bibfield  {journal} {\bibinfo  {journal} {MNRAS}\ }\textbf {\bibinfo
  {volume} {495}},\ \bibinfo {pages} {5027} (\bibinfo {year}
  {2020})}\BibitemShut {NoStop}%
\bibitem [{\citenamefont {Bogadi}\ \emph {et~al.}(2021)\citenamefont {Bogadi},
  \citenamefont {Govender},\ and\ \citenamefont {Moyo}}]{Bogadi2021}%
  \BibitemOpen
  \bibfield  {author} {\bibinfo {author} {\bibfnamefont {R.~S.}\ \bibnamefont
  {Bogadi}}, \bibinfo {author} {\bibfnamefont {M.}~\bibnamefont {Govender}}, \
  and\ \bibinfo {author} {\bibfnamefont {S.}~\bibnamefont {Moyo}},\ }\href
  {\doibase https://doi.org/10.1140/epjc/s10052-021-09744-y} {\bibfield
  {journal} {\bibinfo  {journal} {Eur. Phys. J. C}\ }\textbf {\bibinfo {volume}
  {81}},\ \bibinfo {pages} {922} (\bibinfo {year} {2021})}\BibitemShut
  {NoStop}%
\bibitem [{\citenamefont {Bogadi}\ and\ \citenamefont
  {Govender}(2022)}]{Bogadi2022}%
  \BibitemOpen
  \bibfield  {author} {\bibinfo {author} {\bibfnamefont {R.~S.}\ \bibnamefont
  {Bogadi}}\ and\ \bibinfo {author} {\bibfnamefont {M.}~\bibnamefont
  {Govender}},\ }\href {\doibase
  https://doi.org/10.1140/epjc/s10052-022-10442-6} {\bibfield  {journal}
  {\bibinfo  {journal} {Eur. Phys. J. C}\ }\textbf {\bibinfo {volume} {82}},\
  \bibinfo {pages} {475} (\bibinfo {year} {2022})}\BibitemShut {NoStop}%
\bibitem [{\citenamefont {Alford}\ \emph {et~al.}(2013)\citenamefont {Alford},
  \citenamefont {Han},\ and\ \citenamefont {Prakash}}]{Alford2013}%
  \BibitemOpen
  \bibfield  {author} {\bibinfo {author} {\bibfnamefont {M.~G.}\ \bibnamefont
  {Alford}}, \bibinfo {author} {\bibfnamefont {S.}~\bibnamefont {Han}}, \ and\
  \bibinfo {author} {\bibfnamefont {M.}~\bibnamefont {Prakash}},\ }\href
  {\doibase 10.1103/PhysRevD.88.083013} {\bibfield  {journal} {\bibinfo
  {journal} {Phys. Rev. D}\ }\textbf {\bibinfo {volume} {88}},\ \bibinfo
  {pages} {083013} (\bibinfo {year} {2013})}\BibitemShut {NoStop}%
\bibitem [{\citenamefont {Sennett}\ \emph {et~al.}(2017)\citenamefont {Sennett}
  \emph {et~al.}}]{Sennett2017}%
  \BibitemOpen
  \bibfield  {author} {\bibinfo {author} {\bibfnamefont {N.}~\bibnamefont
  {Sennett}} \emph {et~al.},\ }\href {\doibase 10.1103/PhysRevD.96.024002}
  {\bibfield  {journal} {\bibinfo  {journal} {Phys. Rev. D}\ }\textbf {\bibinfo
  {volume} {96}},\ \bibinfo {pages} {024002} (\bibinfo {year}
  {2017})}\BibitemShut {NoStop}%
\bibitem [{\citenamefont {Romano}(2022)}]{Romano2022}%
  \BibitemOpen
  \bibfield  {author} {\bibinfo {author} {\bibfnamefont {A.~E.}\ \bibnamefont
  {Romano}},\ }\href {https://arxiv.org/abs/2211.05760} {\bibfield  {journal}
  {\bibinfo  {journal} {arXiv:2211.05760 [gr-qc]}\ } (\bibinfo {year}
  {2022})}\BibitemShut {NoStop}%
\end{thebibliography}
\end{document}